\documentclass{llncs}

\usepackage[utf8]{inputenc}
\usepackage[english]{babel}

\usepackage{amsmath,amsfonts,amssymb}
\usepackage{array,tabularx,arydshln,verbatim}
\usepackage{url}
\usepackage{microtype}

\usepackage{tikz,pgfplots}
\usetikzlibrary{calc,shapes.geometric}
\usetikzlibrary{patterns}

\tikzset{
  baseline=(current bounding box.center)
  line cap=round,
  EMline/.style={black!50,dashed},
}

\pgfplotscreateplotcyclelist{mycolor}{%
  red, every mark/.append style={solid,scale=0.8}, mark=x \\%
  blue, every mark/.append style={solid,scale=0.6}, mark=* \\%
  green!70!black, every mark/.append style={solid,scale=0.8}, mark=diamond \\%
  brown!70!black, every mark/.append style={solid,scale=0.6}, mark=square \\%
  teal, every mark/.append style={solid,scale=0.95}, mark=asterisk, dashed \\%
  brown!70!black, every mark/.append style={solid,scale=0.95}, mark=triangle, dashed \\%
  gray, every mark/.append style={solid,scale=0.8}, mark=o, densely dotted \\%
  purple!80!blue, every mark/.append style={solid,scale=0.7}, mark=square, densely dotted \\%
}

\pgfplotsset{
  major grid style={thin,dotted,color=black!50},
  minor grid style={thin,dotted,color=black!50},
  grid,
  cycle list name={mycolor},
  every axis/.append style={
    line width=0.5pt,
    tick style={
      line cap=round,
      thin,
      major tick length=4pt,
      minor tick length=2pt,
    },
  },
  legend cell align=left,
  legend style={
    /tikz/every even column/.append style={column sep=3mm,black},
    /tikz/every odd column/.append style={black},
  },
  every axis label/.append style={font=\small},
  every tick label/.append style={font=\small},
  xlabel near ticks,
  ylabel near ticks,
  plotMini/.style={
    width=63.0mm,height=48mm,
    title style={yshift=-2pt},
  },
  plotMiniBulkSize/.style={
    max space between ticks=18pt,
  },
}

\pgfplotsset{
    ylabel right/.style={
        after end axis/.append code={
            \node [rotate=270, anchor=south, yshift=3pt] at (rel axis cs:1,0.5) {#1};
        }
    }
}

\usepackage{listings}
\AtBeginDocument{%
}

\usepackage{hyperref}
\hypersetup{
  breaklinks=true,
  colorlinks=true,
  citecolor=blue,
  linkcolor=blue,
  urlcolor=blue,
  bookmarksnumbered,
  bookmarksopen,
  pdftitle={A Bulk-Parallel Priority Queue in External Memory with STXXL},
  pdfauthor={Timo Bingmann, Thomas Keh, Peter Sanders},
  pdfsubject={bulk-parallel priority queue, external memory priority queue, stxxl},
  pdfkeywords={bulk-parallel priority queue, external memory priority queue, stxxl},
}

\hypersetup{pdfpagescrop={92 62 523 748}}

\usepackage{breakurl}

\newcommand{\Oh}[1]{\mathcal{O}\!\left( #1\right)}

\newcommand{\Th}[1]{\Theta\!\left( #1\right)}

\title{A Bulk-Parallel Priority Queue \texorpdfstring{\\}{} in External Memory with STXXL}
\author{Timo Bingmann, Thomas Keh, and Peter Sanders}
\institute{
Karlsruhe Institute of Technology, Karlsruhe, Germany\\
\email{\{bingmann,sanders\}@kit.edu}}

\def\clap#1{\hbox to 0pt{\hss#1\hss}}

\begin{document}

\maketitle

\begin{abstract}
  We propose the design and an implementation of a bulk-parallel external memory
  priority queue to take advantage of both shared-memory parallelism and high
  external memory transfer speeds to parallel disks. To achieve higher
  performance by decoupling item insertions and extractions, we offer two
  parallelization interfaces: one using ``bulk'' sequences, the other by
  defining ``limit'' items. In the design, we discuss how to parallelize
  insertions using multiple heaps, and how to calculate a dynamic prediction
  sequence to prefetch blocks and apply parallel multiway merge for
  extraction. Our experimental results show that in the selected benchmarks the
  priority queue reaches 64\% of the full parallel I/O bandwidth of SSDs and
  49\% of rotational disks, or the speed of sorting in external memory when
  bounded by computation.
\end{abstract}

% TODO: remove this prior to FINAL submission
\pagestyle{plain}

\section{Introduction}

Priority queues (PQs) are fundamental data structures which have numerous
applications like job scheduling, graph algorithms, time forward
processing~\cite{chiang1995external}, discrete event simulation, and many greedy
algorithms or heuristics. They manage a dynamic set of items, and support
operations for inserting new items (\emph{push}), and reading and deleting
(\emph{top}/\emph{pop}) the item smallest w.r.t. some order.

Since the performance of such applications usually heavily depends on the PQ, it
is unavoidable to consider parallelized variants of PQs as parallelism is today
the only way to get further performance out of Moore's law.  However, even the
basic semantics of a parallel priority queue (PPQ) are unclear, since PQ
operations inherently sequentialize and synchronize algorithms. Researchers have
previously focused on parallelizing \emph{main memory} PQs which provide
lock-free concurrent access, and/or relaxed operations delivering \emph{some}
small item.

In this work we propose a PPQ for applications where data \emph{does not} fit
into internal memory and thus requires efficient \emph{external memory
  techniques}. Parallelizing external memory algorithms is one of the main
algorithmic challenges termed as ``Big Data''. We propose a ``bulk'' and a
``limit'' parallelization interface for PQs, since the requirements of external
memory applications are different from those working on smaller PQ instances.
One application of these interfaces is \emph{bulk-parallel time forward
  processing}, where one uses the graph's structure to identify layers of nodes
that can be processed independently. For example, the inducing process of an
external memory suffix sorting algorithm~\cite{bingmann2013inducing} follows
this pattern. This paper continues work started in Thomas Keh's bachelor
thesis~\cite{keh2014bulkparallel}.

We implemented our PPQ design in C++ with OpenMP and
STXXL~\cite{dementiev2008stxxl}, and compare it using four benchmarks against the
fastest EM priority queue implementations available. In our experiments we
achieve 49\% of the full I/O throughput of parallel rotational disks and 64\% of four
parallel solid-state-disks (SSDs) with about 2.0/1.6\,GiB/s read/write performance. We reach these percentages in all experiments except when internal work is clearly the limitation, where our
PPQ performs equally well as a highly tuned sorter. For smaller bulk sequences,
the PPQ's performance gradually degrades, however, already for bulks larger than
20\,K or 80\,K 64-bit integers (depending on the platform) our PPQ outperforms
the best existing parallelized external memory PQ.

After preliminaries and related work, we discuss our parallelization interfaces
in Section~\ref{sec:interfaces}. Central is our PPQ design in
Section~\ref{sec:design} where we deal with parallel insertion and
extraction. Details of our implementation, the rationale of our experiments, and
their results are discussed in Section~\ref{sec:experiments}.

% ------------------------------------------------------------------------------
\subsection{Preliminaries}

A PQ is a data structure holding a set of items, which can be
ordered w.r.t some relation. All PQs support two operations: \emph{insert} or
\emph{push} to add an item, and \emph{deleteMin} or \emph{top} and
\emph{pop} to retrieve and (optionally) remove the smallest item from the
set. In this paper we use the \emph{push}, \emph{top}, \emph{pop}
notation, since our implementation's interface aims to be compatible to the C++
Standard Template Library (STL). Addressable PQs additionally provide a
\emph{decreaseKey} operation, which most notably is used by Dijkstra's shortest
path algorithm; but we omit this function since it is difficult
to provide efficiently in external memory.

We use the external memory (EM) model \cite{vitter1994algorithms}, which assumes
an internal memory (called RAM) containing up to $M$ items, and $D$ disks containing space
for $N$ items, used for input, output and temporary data. Transfer of $B$ items
between disks and internal memory costs one I/O operation, whereas internal
computation is free.  While the EM model is good to describe asymptotically
optimal I/O efficient algorithms, omitting computation time makes the model less
and less practical as I/O throughput increases. For example, data transfer to a
single modern SSD reaches more than 450\,MiB/s (MiB = $2^{20}$ bytes), while sorting
1\,GiB of random 64-bit integers sequentially reaches only about 85\,MiB/s on a current
machine.

Thus exploiting parallelism in modern machines is unavoidable to achieve good
performance with I/O efficient algorithms. For this experimental paper, we
assume a shared memory system with $p$ processors or threads, which have a simple set
of explicit synchronization primitives. In future, one could consider a detailed
theoretical analysis using the parallel external memory
model~\cite{arge2008fundamental}.

% ------------------------------------------------------------------------------
\subsection{Related Work}\label{sec:related}

Much work has already been done on incorporating parallelism into PQs for
internal memory. Very different approaches have emerged from it, and we only
present a few here, since our external memory setting is different.

In the 1990s, many PPQs were developed for the PRAM.  Pinotti and Pucci proposed
$n$-bandwidth-heaps \cite{pinotti1991parallel}, which store $n$ sorted items in a heap
node, and allow bulk insertion and extraction of $n$ items by sorting and merging using an
$n$-processor CREW-PRAM.  Similarly, Deo and Prasad describe a parallel
heap~\cite{deo1992parallel}, which allows insertion and extractions of $\Th{n}$
items with $n$ processors by using more advanced PRAM algorithms.  Brodal et
al. \cite{brodal1998parallel} extend these ideas to allow \emph{decreaseKey}
of $m$ arbitrary elements in $\Oh{n}$ time and $\Oh{m \log n}$ work on an
EREW-PRAM.  Sanders~\cite{sanders1998randomized} developed a randomized PPQ for
$p$ distributed memory machines, where each processor keeps a set of local
elements and insertions are randomly distributed among the processors. The PPQ's
\emph{extractMin} operation retrieves the $p$ globally smallest elements, one
per processor, using an exchange algorithm with probabilistic time guarantees.

In the year 2000 and later, many researchers focused on concurrent PQs
\cite{shavit2000skiplist,sundell2003fast,liu2012lock,linden2013skiplist}, which
aim to synchronize frequent access to a common data structure. Each processor
usually wishes to extracts only a single item, as often needed during task
scheduling, and the main goals are to guarantee fairness and avoiding
starvation of processors. The developed PPQs are mostly based on skip lists and
use expensive atomic operations to allow concurrent access to the data structure
without any locks.

Recently, interest has arisen in \emph{relaxed} PPQs
\cite{henzinger2013quantitative,rihani2014multiqueues,alistarh2014spraylist},
since the performance of concurrent PQs does not scale with the higher number of
processors available in newer machines. Instead of returning \emph{the} smallest
item(s), relaxed PPQs return \emph{some} smallest items, and the quality of the
relaxed PPQs is measured by both performance and the introduced errors.
Bulk-parallel PQs can be viewed as synchronous relaxed PQs
with simple and clear semantics.

External memory PQs are a well-established field, and one can choose from
different I/O optimal designs. The older theoretical designs
\cite{arge2003buffer} involve complex buffering of insertion and deletion to
reach optimal $\mathcal{O}((1/B) \log_{M/B} N/B)$ amortized I/O complexity, and
the hidden constants are high. By using buffered multiway merging of
pre-sorted EM lists~\cite{brodal1998worst}, the theoretical algorithms were soon
simplified. In 1999, Brengel et al. \cite{brengel2000experimental} carried out
an experimental study of PQs in EM that resulted in two very practical
external memory PQ designs.

First, they adapted a \emph{radix heap} for external memory. However, the resulting
monotone PQ's I/O complexity depends on the radix and key universe, and is
usually higher than optimal.  Their second approach is called an \emph{external
  array-heap}. It consists of an internal memory heap and a set of sorted arrays
in external memory. The arrays have a fixed size and are arranged in slots,
assigned to a level. The heap can be viewed as the lowest level. Insert
operations go to the lowest level and overflows in one level cause a transfer
into the next higher level after sorting and merging as necessary.

Sanders \cite{sanders00fast} followed a similar approach and improved it among
other thing by paying much more attention to cache efficiency. The data
structure is called a \emph{sequence heap}. Here, the sorted external arrays are
organized in groups of size~$k$, with $k=\Oh{M/B}$ being chosen small enough that merging
all members of a group will be cache-efficient using $k$-way-merge. Similar to
Brengel's approach, an overflow in one group (respective level) causes the
creation of a larger array in the next group. All groups are connected by an
$R$-way-merger, where $R$ is the number of groups. This PQ
design was implemented for external memory in STXXL~\cite{dementiev2008stxxl},
and later also in TPIE~\cite{petersen2007external}, so it is probably the most
widely used today.

The only previous attempt at parallelizing a PQ for EM, that we could find in
literature, was done in conjunction with a study of asynchronous pipelining in
STXXL~\cite{beckmann2009building}.  They partially parallelized the sequence
heap without touching the sequential PQ semantics. However, this gives only little
opportunity for parallelization -- mostly for merging in groups with large
external arrays.

The most sophisticated parallelization tool we use in our PPQ is the parallel
$k$-way merge algorithm first proposed by Varman et
al. \cite{varman1991merging}, and engineered by Singler et al. in the
MCSTL~\cite{singler2007mcstl} and later the GNU Parallel Mode
library~\cite{singler2007mcstl}.  Since this algorithm's details and
implementation are important for our PPQ design, we briefly describe it: given
$p$ processors and $k$ sorted arrays with in total $n$ items and of maximum
length $m$, each array is split into $p$ range-disjoint parts where the sum of
each processor's parts are of equal size. The partition is calculated by running
$p$ intertwined multisequence selection algorithms, which take
$\mathcal{O}(k \cdot \log k \cdot \log m)$. After partitioning, the work of
merging the $p$ disjoint areas can be done independently by the
processors, e.g., using a $k$-way tournament tree in time
$\mathcal{O}(\frac{n}{p} \log k)$. For our EM setting it is important that the
output is generated as $p$ equal-sized parts in parallel, with each part being written in sequence. We also note that the multisequence
selection is implemented sequentially.

% ------------------------------------------------------------------------------
\section{Bulk-Parallel Interface and Limit Items}\label{sec:interfaces}

Before we discuss our PPQ design, we focus on the proposed application
interface. As suggested by the related work on PQs, substantial performance
gains from parallelization are only achievable when loosening some semantics of
the PQ. Put plainly, an alternating sequence of dependent \emph{push}/\emph{pop}\,s
is inherently sequential. Since we focus on large amounts of data, the more natural
relaxation of a PQ is to require insertion and extraction of multiple items, or
``bulks'' of items. This looser semantic \emph{decouples} insert and delete
operations both among themselves (i.e., items within a bulk) as well as the
operation phases from another. This enables us to apply parallel algorithms on
larger amounts of items, and our experiments in Section~\ref{sec:experiments}
show how speedup depends on the bulk sizes.

Thus the primary interface of our EM PPQ is bulk insertion and extraction (see
Listing~\ref{lst:bulk-pop-push}). A bulk insertion phase is started with
\emph{bulk\_push\_begin$(k)$}, where $k$ is an estimate of the bulk size, which we
will use to optimize preparation in the PPQ. Thereafter, the application may
insert a bulk of items using \emph{bulk\_push}, possibly concurrently from
multiple threads, and terminate the sequence with \emph{bulk\_push\_end}. There
are two bulk extraction primitives: \emph{bulk\_pop$(v,k)$} which extracts up to
$k$ items into $v$, and \emph{bulk\_pop\_limit$(v,L,k)$}, which extracts at most
$k$ items strictly smaller than a limit item $L$. The limit extraction also indicates
whether more items smaller than $L$ are available.

\begin{figure}[t]
\begin{minipage}[t]{0.54\linewidth}
\begin{lstlisting}[caption={Bulk Pop\smash{/}Push Loop},frame=single,label=lst:bulk-pop-push]
vector<item> work;
while (!ppq.empty()) {
  ppq.bulk_pop(work, max_size);
  ppq.bulk_push_begin(approx_bulk_size);
#pragma omp parallel for
  for (i = 0; i < work.size(); ++i) {
    // process work[i], maybe bulk_push()
  }
  ppq.bulk_push_end();
}
\end{lstlisting}
\end{minipage}
\hfill%
\begin{minipage}[t]{0.42\linewidth}
\begin{lstlisting}[caption={Bulk-Limit Loop},frame=single,label=lst:bulk-limit]
for (...) {
  ppq.limit_begin(L, bulk_size);
  while (ppq.limit_top() < L) {
    top = ppq.limit_top();
    ppq.limit_pop();
    // maybe use limit_push()
  }
  ppq.limit_end();
}
\end{lstlisting}
\end{minipage}
\vspace{-3ex}
\end{figure}

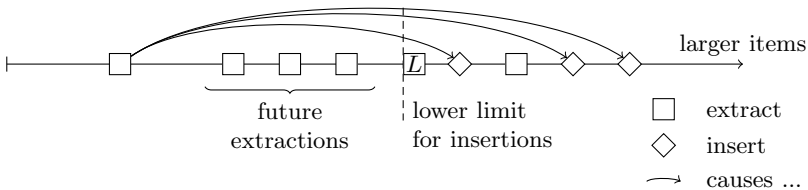
\begin{figure}[b]\centering
  \usetikzlibrary{shapes,decorations}
  \begin{tikzpicture}[
    xscale=1.5,
    rdot/.style={draw,rectangle,inner sep=0pt,minimum size=8pt,fill=white},
    ddot/.style={draw,diamond,inner sep=0pt,minimum size=9pt,fill=white},
    decoration=brace,
    looseness=0.7,
    ]

    \draw[->] (0,0) -- (6.5,0) node [above] {larger items};
    \draw (0,0.1) -- (0,-0.1);

    \node[rdot] (r0) at (1,0) {};
    \node[rdot] (r1) at (2,0) {};
    \node[rdot] (r2) at (2.5,0) {};
    \node[rdot] (r3) at (3,0) {};

    \draw[densely dashed] (3.5,0.75) -- (3.5,-0.75);
    \node[rdot,anchor=west] (r3) at (3.5,0) {$L$};

    \node[ddot] (d0) at (4,0) {};
    \node[rdot] (r4) at (4.5,0) {};
    \node[ddot] (d1) at (5,0) {};
    \node[ddot] (d2) at (5.5,0) {};

    \path (2,0.6) -- ++(1,0);

    \draw[overlay] (r0) edge [->,out=45,in=180-45]  (d0);
    \draw[overlay] (r0) edge [->,out=45,in=180-45]  (d1);
    \draw[overlay] (r0) edge [->,out=45,in=180-45]  (d2);

    \node[align=left] at (4.2,-0.8) {lower limit\\for insertions};
    \draw[decorate,shift={(1.75,-0.3)}]  (1.5,0) -- node[align=center,below=0.7ex] {future\\extractions} (0,0);

    \begin{scope}[yshift=-6mm,xshift=58mm,yscale=0.8]
      \node[rdot] at (0,0) {};
      \node[right] at (0.3,0) {extract};

      \node[ddot] at (0,-0.6) {};
      \node[right] at (0.3,-0.6) {insert};

      \draw (-0.15,-1.2) edge [->,out=60,in=180-45] ++(0.3,0);
      \node[right] at (0.3,-1.2) {causes ...};
    \end{scope}

  \end{tikzpicture}
  \caption{Decoupling insertion and extraction operations with a limit item $L$.}\label{fig:limit bulk}
\end{figure}

Beyond the primary bulk interface, we also propose a second interface (see
Listing~\ref{lst:bulk-limit}), which is geared towards the canonical processing
loop found in most sequential applications using a PQ: extract an item, inspect
it, and reinsert zero or more items into the PQ. To decouple insertions and
extractions in this loop, we let the application define a ``limit item'' $L$, and
require that all insertions thereafter are larger or equal to $L$ (see
Figure~\ref{fig:limit bulk}). By defining this limit, all extractions of items
less than $L$ become decoupled from insertions. The drawback of this second
interface is that the application does not process items in parallel.  However,
parallel processing of items $< L$ can easily be accomplished by using
\emph{bulk\_pop\_limit} in the Bulk Pop/Push Loop example.

% ------------------------------------------------------------------------------
\section{Design of a Bulk-Parallel Priority Queue}\label{sec:design}

\begin{figure}[t]\centering
\begin{tikzpicture}[scale=1.6,
  IM/.style={fill=black!10},
  EM/.style={pattern=north west lines},
  minimatree/.style={
    draw, IM, isosceles triangle,rotate=90, inner sep=0,
    minimum size=3mm, isosceles triangle apex angle=70
  },
  ]

  \def\heap at (#1,#2,#3){
    \draw[IM] (#1,#2+1) -- (#1+0.65,#2) -- (#1-0.65,#2) -- (#1,#2+1);
    \coordinate (insheap#3) at (#1,#2+1)
  }

  \begin{scope}[scale=0.4]
    \heap at (-0.9,0,1);
    \heap at (+0.9,0,2);
    \heap at (-0.9,1.5,3);
    \heap at (+0.9,1.5,4);
  \end{scope}

  \node[align=center,anchor=base] at (0,-0.6) {$p$ insertion\\heaps};

  \node (MT1) at (0,1.4) [minimatree] {};

  \draw (insheap3) to[out=90,in=-90] (MT1.{180-50});
  \draw (insheap1) to[out=90,in=-90] ++(0.31,0.2) -- (MT1.{180-25});
  \draw (insheap2) to[out=90,in=-90] ++(-0.31,0.2) -- (MT1.{180+25});
  \draw (insheap4) to[out=90,in=-90] (MT1.{180+50});

  \begin{scope}[xshift=11mm]

    \draw[IM] (0.0,0) rectangle ++(0.2,1);
    \draw[IM] (0.4,0) rectangle ++(0.2,1);
    \node at (0.9,0.5) {...};
    \draw[IM] (1.2,0) rectangle ++(0.2,1);

    \node[rotate=270] at (0.1,0.5) {$<$};
    \node[rotate=270] at (0.5,0.5) {$<$};
    \node[rotate=270] at (1.3,0.5) {$<$};

    \node[align=center,anchor=base] at (0.7,-0.6) {internal\\arrays};

    \node (MT2) at (0.7,1.4) [minimatree] {};
    \draw (0.1,1) to[out=90,in=-90] (MT2.{180-50});
    \draw (0.5,1) to[out=90,in=-90] (MT2.{180-0});
    \draw (1.3,1) to[out=90,in=-90] (MT2.{180+50});

  \end{scope}

  \begin{scope}[xshift=34mm]

    \draw[IM] (-0.15,0) rectangle (0.15,1);
    \node[align=center,anchor=base] at (0,-0.6) {extract\\buffer};

    \node[rotate=270] at (0,0.5) {$<$};

    \coordinate (EB) at (0,1);

  \end{scope}

  \begin{scope}[xshift=42mm]

    \def\ea at (#1,#2){
      \draw[EM] (#1,#2 + 0.1) -- ++(0,0.9) -- ++(0.2,0) -- ++(0,-0.9) to[out=135,in=310] (#1,#2 + 0.1);
      \draw[densely dotted] (#1,#2 + 0.1) -- ++(0,-0.2);
      \draw[densely dotted] (#1 + 0.2,#2 + 0.1) -- ++(0,-0.2);
      \draw[IM] (#1,#2 + 1.0) rectangle ++(0.2,-0.2);
    };

    \ea at (0.0,0);
    \ea at (0.4,0);
    \node at (0.9,0.5) {...};
    \ea at (1.2,0);

    \node[align=center,anchor=base] at (0.7,-0.6) {external\\arrays};

    % \node (MT3) at (0.7,1.4) [minimatree] {};
    % \draw (0.1,1) to[out=90,in=-90] (MT3.{180-50});
    % \draw (0.5,1) to[out=90,in=-90] (MT3.{180-0});
    % \draw (1.3,1) to[out=90,in=-90] (MT3.{180+50});

  \end{scope}

  \begin{scope}[xshift=61mm,yshift=-3mm]

    \foreach \x in {0,1,2,3,4} {
      \draw[IM] (-0.15,\x * 0.2) rectangle ++(0.2,0.2);
    }
    \node[align=center,anchor=base,rotate=90] at (0.3,0.5) {read/write buffers};

  \end{scope}

  \begin{scope}[looseness=0.3]
    \node (MT) at (2.5,2.0) [minimatree] {};

    \draw (MT1.0) to[out=90,in=-90] (MT.{180-50});
    \draw (MT2.0) to[out=90,in=-90,looseness=0.9] (MT.{180});
    \draw (EB)    to[out=90,in=-90,looseness=0.9] (MT.{180+50});
    %\draw (MT3.0) to[out=90,in=-90] (MT.{180+50});
  \end{scope}

  \begin{scope}[densely dotted, semithick]
    %\draw (-0.8,-0.8) rectangle ++(1.6,2.0);
    %\draw (0.8,-0.8) rectangle ++(1.4,2.0);
    %\draw (2.2,-0.8) rectangle ++(1.8,2.0);
    %\draw (4.0,-0.8) rectangle ++(1.8,2.0);
    %\draw (5.8,-0.8) rectangle ++(0.7,2.0);
    %\draw (-0.8,1.2) rectangle ++(7.3,1.0);
    \draw (-0.8,-0.8) rectangle (6.6,2.2);
    \draw (-0.8,1.2) -- (6.6,1.2);
    \draw (0.8,-0.8) -- ++(0,2);
    \draw (2.8,-0.8) -- ++(0,2);
    \draw (4.0,-0.8) -- ++(0,2);
    \draw (5.8,-0.8) -- ++(0,2);
    \node at (4.2,1.9) {hierarchy of tournament trees};
  \end{scope}

\end{tikzpicture}
\caption{Components and architecture of our PPQ. All lightly shaded parts are in internal memory, diagonally lined ones in EM.}\label{fig:architecture}
\end{figure}
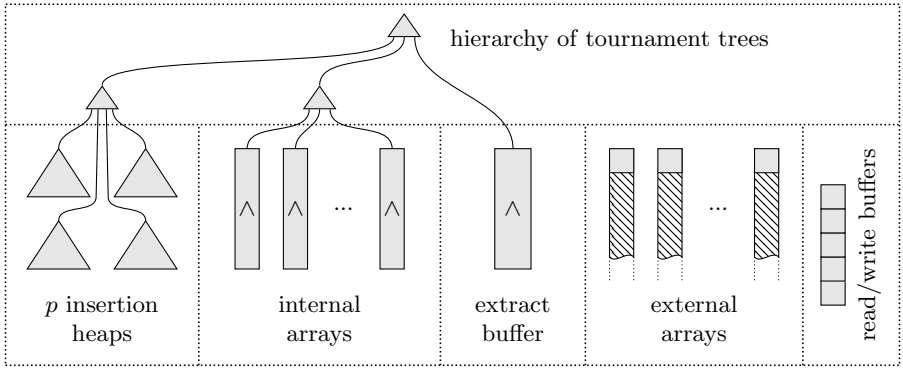

Our PPQ design (see Figure~\ref{fig:architecture}) is based on Sanders' sequence
heap \cite{sanders00fast}, but we have to reevaluate the implicit assumptions,
duplicate data structures for independent parallel operations and apply parallel
algorithms where possible. After briefly following the lifetime of an item in
the PPQ, we first discuss separately how insertions and extractions can be
processed in parallel, and then focus on the difficulty of balancing both.

An item is first inserted into an \emph{insertion heap}, which is kept
in heap order. As simple binary heaps are not particularly cache-efficient, they are given a
fixed maximum size. When full, an insertion heap is sorted and transformed into
an \emph{internal array}. To limit the number of internal arrays, they may be
merged with others to form longer internal arrays. When memory is exhausted, all
internal arrays and the extract buffer are merged into one sorted \emph{external
  array} which is written to disk. Again, shorter external arrays may also be
merged together. Extracts from the set of external arrays are amortized using
the extract buffer.

\textbf{Insertion, Multilevel Merging, and External Writing.}
To accelerate parallel push operations, the first obvious step is to have $p$
insertion heaps, one for each processor. This decouples insertions on different
processors and parallelizes the work of maintaining the heaps. Once a heap is
full, the processor can independently sort the heap using a general sorter (heap
sort is usually slower). Remarkably, these initial steps are among the most time
consuming in a sequence heap, and can be parallelized well. In our PPQ
design, we then use a critical section primitive to synchronize adding the new
internal array to the common list. This was never a bottleneck, since such
operations happen only when an insertion heap is full; however, one could also
use a lock-free queue or array to boost performance.

In bulk push sequences, we can accelerate individual push operations much further. While pushing, no items from the insertion heap can be extracted, thus we can
postpone reestablishing heap order to \emph{bulk\_push\_end}; a
\emph{bulk\_push} just appends to the insertion heap's array. If the heap
overflows, then the array is sorted anyway. In our experiments, this turned out
to be the best option, probably because the loop sifting items up the heap
becomes very tight and cache efficient. For larger bulk operations (as indicated
by the user's estimation) we even let the insertion heap's array grow beyond the
usual limit to fill up the available RAM, since sorting is more cache efficient
than keeping a heap.

Instead of separating internal arrays into groups, as in a sequence heap,
we label them using a \emph{level number} starting at zero. If the number of
internal arrays on one level grows larger than a tuning parameter (about
64) and there is enough
RAM available, then all internal arrays of one level are merged together and
added to the next higher level.

The decisive difference of parallel multiway merge over sequentially merging
sorted arrays is that \emph{no state} is kept to amortize operations. Hence, in
our PPQ design the indicated tournament trees over the insertion heaps and
arrays are useless for parallel operations. When applying parallel multiway
merge, we want to have the total number of items as large as possible, however,
at the same time the number of sequences should be kept as small as possible,
especially since the multisequence selection is currently implemented
sequentially. In Appendix~\ref{sec:multiwaymerge} we report on preliminary
experiments to find a good balance.

When the PPQ's alloted memory is exhausted, mostly due to long internal arrays, one
large parallel multiway merge is performed directly into EM. This is possible
without an extra copy buffer, by using just $\Theta(p)$ write buffers and
overlapping I/O and computation, since parallel multiway merge outputs $p$
sorted sub-sequences. We use $\geq 2 p$ write buffer blocks to keep the merge
boundaries in memory; thus avoiding any rereading of blocks from disk during
the merge. Another method would be to round the ranks during multisequence
selection to block boundaries. This would slightly disbalance work, but removes all access conflicts within blocks.

An item may travel multiple times to disk and back, since the extract buffer is
included while merging into EM. However, as in the sequence heap structure, this
only occurs when internal memory is exhausted and all items are written to disk;
thus we can amortize the extra I/Os for the extract buffer with the
$\Theta(M/B)$ I/Os needed to flush main memory.

\textbf{Extraction, Prediction, and Minimum of Minima.}
To support fast non-bulk \emph{pop} operations, we keep a hierarchy of
tournament trees to save results of pairwise comparisons of items. The trees are
built over the insertion heaps, internal arrays, and extract buffer. External
arrays need not be included, since extraction from them is buffered using the
extract buffer. The tournament trees need to be updated each time an insertion
heap's minimum element changes, or a heap is flushed into an internal array. In
bulk push operations these actions are obviously postponed until the bulk's end.

\begin{figure}[t]
  \begin{tikzpicture}[scale=1,
    inner sep=0, outer sep=0,
    IM/.style={fill=black!10},
    FINISHED/.style={preaction={},pattern=crosshatch, pattern color=black},
    FINISHEDBEGIN/.style={preaction={},pattern=crosshatch, pattern color=black},
    EM/.style={pattern=north west lines,preaction={}},
    EMEND/.style={pattern=north west lines,preaction={}},
    HINTED/.style={pattern=north west lines,preaction={fill, black!30}, pattern color=black!80},
    HINTEND/.style={pattern=north west lines,preaction={fill, black!30}, pattern color=black!80},
    HEAP/.style={fill=black!10},
    ]

    \pgfmathsetmacro{\size}{0.4}

    \def\heap at (#1,#2,#3){
      \draw[HEAP] (#1+0.7,#2) -- (#1+1.4,#2-1) -- (#1,#2-1) -- (#1+0.7,#2);
      \coordinate (heap#3) at (#1,#2)
    }

    \def\minkey[#1] at (#2,#3){
      \node[align=center,anchor=south] at (#2+0.05,#3+0.4*\size) {\scriptsize #1};
    };

    \def\heapkey[#1] at (#2,#3,#4){
      \node[align=left,anchor=base, outer sep=2pt, inner sep=1pt] at (#2+0.3+#4,#3-1.25) {\scriptsize #1};
    };

    \def\heaptitle[#1] at (#2,#3){
      \fontsize{9}{9}\selectfont
      \node[align=center,anchor=north, text width=2.6cm] at (#2+0.7,#3+1.2) {#1};
    };

    \def\block[#1] at (#2,#3){
      \draw[#1] (#2,#3) rectangle ++(\size,-\size);
      % minimum
      \draw[IM] (#2,#3) rectangle ++(\size/4,\size/4);
    };

    \pgfmathsetmacro{\legendsize}{\size/1.3}

    \def\legendblock[#1] at (#2,#3) (#4){
      \draw[#1,draw=black] (#2,#3) rectangle ++(\legendsize,-\legendsize);
      \fontsize{9pt}{9}\selectfont
      \node[align=left,anchor=west] at (#2+\legendsize+0.15,#3-\legendsize/2) {#4};
    };

    \def\emend at (#1,#2){
      \draw[EMEND] (#1,#2) -- ++(\size,0) to[out=-135,in=45] ++(0,-\size) --  (#1,#2-\size);
      \draw[densely dotted] (#1+\size,#2) -- ++(\size/2,0);
      \draw[densely dotted] (#1+\size,#2-\size) -- ++(\size/2,0);
      % minimum
      \draw[IM] (#1,#2) rectangle ++(\size/4,\size/4);
    };

    \def\finishedbegin at (#1,#2){
      \draw[FINISHEDBEGIN] (#1+\size,#2-\size) -- ++(-\size,0) to[out=45,in=-135]  (#1,#2) -- ++(\size,0);
      \draw[densely dotted] (#1,#2) -- ++(-\size/2,0);
      \draw[densely dotted] (#1,#2-\size) -- ++(-\size/2,0);
      \draw (#1+\size,#2) -- ++(0,-\size);
      % minimum
      % \draw[IM] (#1+\size+\size/4,#2) rectangle ++(\size/4,-\size/4);
    };

    \def\finished at (#1,#2){
      \block[FINISHED] at (#1,#2);
    };

    \def\im at (#1,#2){
      \block[IM] at (#1,#2);
    };

    \def\em at (#1,#2){
      \block[EM] at (#1,#2);
    };

    \def\hinted at (#1,#2){
      \block[HINTED] at (#1,#2);
    };

    \def\hintend at (#1,#2){
      \block[HINTEND] at (#1,#2);
    };

    % winner trees
    \begin{scope}[xshift=-0.2cm,yshift=-1.1cm]
      \begin{scope}[xshift=-2.1cm,yshift=0cm]
        \heap at (-2,0,1);
        \heapkey[$m_0$] at (-2,0,-0.1);
        \heapkey[$m_1$] at (-2,0,0.4);
        \heapkey[$m_2$] at (-2,0,0.9);
        \heaptitle[{tree of next \\ \textit{loadable} \\ minima}] at (-2,-0.2);
      \end{scope}

      \begin{scope}[xshift=0cm,yshift=0cm]
        \heap at (-2,0,1);
        \heapkey[$h_0$] at (-2,0,-0.1);
        \heapkey[\raisebox{0.4pt}{$\infty$}] at (-2,0,0.4);
        \heapkey[$h_2$] at (-2,0,0.9);
        \heaptitle[tree of next \\ \textit{hintable} \\ minima] at (-2,-0.2);
      \end{scope}

    \end{scope}

    % external arrays
    \begin{scope}[xshift=0.2cm]

      \begin{scope}[xshift=0cm,yshift=-0cm]
        \finishedbegin at (0,0);
        \finished at (\size,0);
        \im at (2*\size,0);
        \im at (3*\size,0);
        \im at (4*\size,0);
        \im at (5*\size,0);
        \coordinate (m0) at (5*\size,0);
        \minkey[$m_0$] at (6*\size,0);
        \hinted at (6*\size,0);
        \hinted at (7*\size,0);
        \hinted at (8*\size,0);
        \minkey[$h_0$] at (9*\size,0);
        \emend at (9*\size,0);
      \end{scope}

      % EA1
      \begin{scope}[xshift=0cm,yshift=-1cm]
        \finishedbegin at (0,0);
        \im at (\size,0);
        \im at (2*\size,0);
        \im at (3*\size,0);
        \im at (4*\size,0);
        \minkey[$m_1$] at (5*\size,0);
        \coordinate (m1) at (5*\size,0);
        \hinted at (5*\size,0);
        \hinted at (6*\size,0);
        \hinted at (7*\size,0);
        \hinted at (8*\size,0);
        \hintend at (9*\size,0);
      \end{scope}

      % EA3
      \begin{scope}[xshift=0cm,yshift=-2cm]
        \finishedbegin at (0,0);
        \finished at (\size,0);
        \im at (2*\size,0);
        \im at (3*\size,0);
        \im at (4*\size,0);
        \im at (5*\size,0);
        \im at (6*\size,0);
        \minkey[$m_2$] at (7*\size,0);
        \coordinate (m2) at (7*\size,0);
        \hinted at (7*\size,0);
        \minkey[$h_2$] at (8*\size,0);
        \em at (8*\size,0);
        \emend at (9*\size,0);
      \end{scope}

      % merge limit line
      \draw[rounded corners=2pt]
      ($(m0) + (-2mm,3mm)$) |-
      ($(m1) + (0.5mm,1.5mm)$) |-
      node[anchor=south east,xshift=-1mm] {\scriptsize merge limit}
      ($(m2) + (-13mm,2mm)$) -- ++(0,-7mm);

    \end{scope}

    % legend
    \begin{scope}[xshift=5.0cm,yshift=-0.3cm]

      \legendblock[FINISHED] at (0,0) (finished);
      \legendblock[IM] at (0,-\size-0.1) (in RAM);
      \legendblock[HINTED] at (0,-2*\size-0.2) (prefetched);
      \legendblock[EM] at (0,-3*\size-0.3) (external);

    \end{scope}

  \end{tikzpicture}
\caption{Establishing the dynamic prefetching sequence and upper merge limit.}\label{fig:block minima trees}
\end{figure}
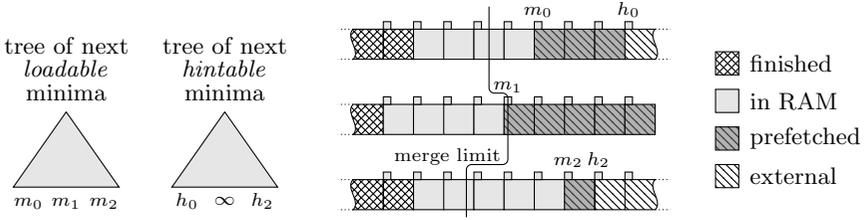

When merging external arrays with parallel multiway merge we are
posed (again) with the discrepancy between parallelism, which requires large
item counts for efficiency, and relatively small disk blocks (by default
2--8\,MiB). To alleviate the problem, we increase the number of read buffers and
calculate an \emph{optimal block prediction sequence}, as also done for
sorting~\cite{hutchinson2005duality}, which contains the order in which the EM blocks are needed
during merging and fetch as many as fit into RAM. In sorting, the prediction
sequence is fixed and can be determined by sorting the smallest items of each
block as a representative (also called ``trigger'' element). In the parallel
disk model, the independent disks need to be considered as well. In our PPQ
setting, the prediction sequence becomes a \emph{dynamic problem}, since
external arrays may be added. We define four states for an external block: in
external memory, hinted for prefetching, loaded in RAM, and finished (see
Figure~\ref{fig:block minima trees}). To limit the main memory usage of the PPQ,
the number of prefetched and blocks loaded in RAM must be restricted.

Since the next $k$ external blocks needed for merging are determined by the $k$
smallest block minima, we keep track of these items in a tournament tree over
the block minima sequences of the external arrays (see items $h_i$ in
Figure~\ref{fig:block minima trees}). This allows fast calculation of the next
block when another can be prefetched. However, when a new external array is
added, the dynamic prediction sequence changes, and we may have to cancel
prefetch hints. This is done by resetting the tournament tree back to the
first block minima merely hinted for prefetching, but not loaded in RAM, and
replaying it till the new $k$ smallest block minima are determined. This costs
less than $k + k \log S$ comparisons, where $S$ is the number of sequences. We
then compare the new predictions with the old ones simply by checking how many
blocks are to be prefetched in each array, and cancel or add prefetch hints.

For parallel merging, however, we need to solve another problem: the merge
ranges within the blocks in RAM must be limited to items smaller than (or equal
to) the smallest item still in EM, since otherwise the PQ invariant
may be violated. To determine the smallest item in EM we reuse the
block minima sequences, and build a second tournament tree over them containing
the smallest items of the next ``loadable'' block, not guaranteed to be in RAM
(items $m_i$ in Figure~\ref{fig:block minima trees}). When performing a parallel
multiway merge into the extract buffer, all hinted external blocks are first
checked (in order) whether the prefetch is complete, and the tournament tree
containing the smallest external items is updated. The tip then contains
$\overline{m} = \min_i m_i$, the overall smallest external item, which serves as \emph{merge
  limit}. We then use binary search within the loaded blocks of each array and find the largest
items smaller than $\overline{m}$, (or if one sacrifices stability of the PPQ, the largest
items smaller or equal to $\overline{m}$; if one defines stability appropriately).

We thus limit the multisequence selection and merge range on each array by
$\overline{m}$. Additionally, by using smaller selection ranks during parallel multiway
merging one can adapt the total number of elements merged. These rank limits
enable us to efficiently limit the extract buffer's size and the output size of
\emph{bulk\_pop$(v,k)$} and \emph{bulk\_pop\_limit$(v,L,k)$} operations. To
limit extraction up to $L$, we simply use $\min(L,\overline{m})$ as merge limit.

As with internal arrays, the number of external arrays should be kept small
for multiway merge to be efficient. One may suspect that merging from EM
is I/O bound, however, if the merge output buffers are smaller than the read
buffers, then this is obviously not the case. Thus, the parallelization
bottleneck of refilling the extract buffer or of \emph{bulk\_pop} operations
largely depends on the number of arrays. We also adapt the number of read
buffers (both for prefetching and holding blocks in RAM) dynamically to the number of
external arrays. Each newly added external array requires at least
one additional read buffer, since otherwise one cannot guarantee that the
first block is loadable if needed.

As with internal arrays, instead of keeping external arrays in separate groups,
we label them with a level number, and merge levels when the contained number
grows too large. This enables more dynamic memory pooling than in the rigid
sequence heap data structure, while maintaining the optimal I/O complexity.

\textbf{Trade-Offs between Insertion and Extraction.}  As already discussed, to
enable non-bulk \emph{pop} operations we keep a hierarchy of tournament
trees. Using this hierarchy instead of one large tournament tree skews the depth
of nodes in the tree, making replays after \emph{pop}s from the extract buffer
and the insertion heaps cheaper than from internal arrays.

When a new external array is created, then the read prediction sequence may
change and previous prefetch requests need to be canceled and new ones
issued. In long bulk push sequences (as the ascending sequence in our
experiments), this can amount to many superfluous prefetch reads of blocks. Thus
we disable prefetching during \emph{bulk\_push} operations and issue all hints
at the end. This suggests that bulk push sequences should be as long as
possible, and that they are interleaved with \emph{bulk\_pop} operations.

% ------------------------------------------------------------------------------
\section{Implementation in STXXL and Experimental Results}\label{sec:experiments}

We implemented our PPQ design in C++ with OpenMP and the STXXL
library~\cite{dementiev2008stxxl}, since it
provides a well-designed interface to asynchronous I/O, and allowing easy overlapping
of I/O and computation. It also contains two other PQ implementations that we
compare our implementation to. Our implementation will be available as part of
the next STXXL release 1.4.2, which will be publicly available under the liberal
Boost software license. At the time of submission it is available in the public development repository.

Concerning actual use of the PPQ, we must point out that contrary to the previous
description in this paper, the implementation extracts the \emph{largest items}
w.r.t. the given order first. This is because the EM containers in STXXL follow
the C++ STL's interface and \texttt{std::priority\_queue} is a max-heap.

\textbf{Other PQ Implementations.}  In these experiments we compare our PPQ
implementation (\textbf{PPQ}) with the sequential sequence
heap~\cite{sanders00fast} (\textbf{SPQS}) contained in the STXXL, a partially
parallelized version~\cite{beckmann2009building} of it (\textbf{SPQP}),
which uses parallel multiway merging only when merging external arrays, and with
the STXXL's highly tuned stream sorting
implementation~\cite{dementiev2003asynchronous} (\textbf{Sorter}) as a baseline.

\textbf{Experimental Platforms.}  We run the experiments on two platforms (see
also Table~\ref{tbl:platforms} in the appendix). Platform \textbf{A-Rot} is an
Intel Xeon X5550 from 2009 with 2 sockets, 4 cores and 4 Hyperthreading cores
per socket at 2.66\,GHz clock speed and 48\,GiB RAM, and eight rotational Western Digital Blue
disks with 1\,TB capacity and about 127\,MiB/s transfer speed each, which
are attached via an Adaptec ASR-5805 RAID controller. Platform \textbf{B-SSD} is
an Intel Xenon E5-2650~v2 from 2014 with 2 sockets, 8 cores and 8 Hyperthreading
cores per socket at 2.6\,GHz clock speed with 128\,GiB RAM. There are four
Samsung SSD 840 EVO disks with 1\,TB each attached via an Adaptec ASA-7805H Host
adapter, yielding together 2\,GiB/s read and 1.6\,GiB/s write transfer speed
to/from EM. The platforms run Ubuntu Linux 12.04 and 14.04, respectively, and
all our programs were compiled with gcc 4.6.4 and 4.8.2 in \emph{Release}
performance mode using STXXL's CMake build system.

\textbf{Experiments and Parameters.} To compare the three PQs we report results
of four sets of experiments. In all experiments the PQ's items are plain 64-bit
integer keys (8~bytes), which places the spotlight on internal comparison work
as payload only increases I/O volume. (See Figure~\ref{fig:results-24} and
Tables~\ref{tab:results-24}--\ref{tab:speedups-24} in the appendix for additional
results with 24\,byte items.) The PQs are allotted 16\,GiB of RAM on both
platforms, since in a real EM application multiple data structures exist
simultaneously and thus have to share RAM.

In the first two experiments, called a) \emph{push-rand-pop} and b)
\emph{push-asc-pop}, the PQ is filled a) with $n$ uniformly random generated
integer items, or b) with $n$ ascending integers, and then the $n$ items are
extracted again. In these canonical benchmarks, the PQ is used to just sort
the items, but it enables us to compare the PQs against the highly optimized
sorting implementation, which also employs parallelism where possible. In the
ascending sequence, the first items inserted are removed first, forcing the PQs
to cycle items. Considering the amount of internal sorting and merging work, the
\emph{push-asc-pop} benchmark is an easy case, since all buffers are
sorted and merging is skewed. Thus the focus of this benchmark is on I/O overlapping. On the other hand, in the
\emph{push-rand-pop} benchmark the internal work to sort and merge the
random numbers is very high, which makes it a test of internal processing
speed. We ran the experiments for $n = 2^{27}, \ldots,2^{35}$, which is an item
volume of $1\text{\,GiB}, \ldots, 256\text{\,GiB}$.

The third and forth experiments, \emph{asc-rbulk-rewrite} and
\emph{bulk-rewrite}, fully rewrite the PQ in bulks: the PQ is filled with $n$
ascending items, then the $n$ items are extracted in bulks of random or fixed
size $v$, and after each bulk extraction $v$ items are pushed again. During
the rewrite, in total $n$ items are extracted and $n$ items inserted with higher
ids. We measure only the bulk pop/push cycles as these experiments are designed
to emulate traversing a graph for time forward processing.  We use bulk
rewriting in two different experiment scenarios: in the first, we select the bulk size
uniformly at random from 0 to 640\,000, and let $n$ increase as in
the first two experiments.  For the second, $n = 4 \cdot 2^{30}\,\text{items}$
(32\,GiB) is fixed and the bulk size $v$ is varied from 5\,000 to 5\,120\,000.

All experiments were run only once due to long execution times and little
variation in the results over large ranges of input size.
During the runs we
pinned the OpenMP threads to cores, which is important since it keeps the
insertion heaps local. Due to the large I/O bandwidth of the SSDs, we increased
the number of write buffers of the PPQ to 2\,GiB on B-SDD to better overlap I/O
and computation. Likewise, we allotted 128\,MiB read buffers per external
array. On A-Rot we set only 256\,MiB write buffers and 32\,MiB read buffers per
array. For the STXXL PQ, of the 16\,GiB of RAM one fourth is allocated for read
and one fourth for write buffers. We used in all experiments the new
``linuxaio'' I/O interface of STXXL 1.4.1, which uses direct system calls to
Linux's asynchronous I/O interface with native command queuing
(NCQ) and bypasses system disk cache.

While STXXL bypasses the system disk cache, we did not disable the write
cache/buffering inside the disks themselves. Disabling these features prohibits
the disks from doing request reordering and asynchronous operation
scheduling. On the SSDs the performance reduction would have been around 20\%,
on the rotation disks about 10\%.

\textbf{Results and Interpretation.}  The results measured in our experiments
are shown in Figure~\ref{fig:results-8} as throughput in items per second, and in
MiB/s in Table~\ref{tab:results-8} in the appendix. We measured ``throughput''
at the PQ interface, and this is not necessarily the I/O throughput to/from
disk, since the PQs may keep items in RAM. In all four experiments, items are
read or written \emph{twice}, so throughput is two times item size divided by
time. If one assumes that a container writes and reads all items once to/from
disk (as the sorter does), then on A-Rot at most 39 million items/s and on
B-SSD at most 106 million item/s could be processed, considering the maximum I/O
bandwidth as measured using \texttt{stxxl\_tool}.

% IMPORT-DATA pqorder plotdata/pqorder.txt
% IMPORT-DATA oporder plotdata/oporder.txt

% IMPORT-DATA stats125 plotdata/20150330/i10pc125-8b/*/size-*.txt
% IMPORT-DATA stats129 plotdata/20150330/i10pc129-8b/*/size-*.txt

% IMPORT-DATA stats125b plotdata/20150330/i10pc125-8b/*/bulk-*.txt
% IMPORT-DATA stats129b plotdata/20150330/i10pc129-8b/*/bulk-*.txt

\begin{table}[b]\centering
  \def\tabcolsep{6pt}
  \begin{tabular}{|r|rrr|rrr|}
    \hline
               & \multicolumn{3}{c|}{Platform A-Rot} & \multicolumn{3}{c|}{Platform B-SSD}  \\
    Experiment & SPQP                                & SPQS & Sorter & SPQP & SPQS & Sorter \\ \hline
    %% TABULAR REFORMAT(precision=2)
    %% SELECT opo.desc,
    %% MEDIAN(spq125.time / ppq125.time) AS spq_ratio,
    %% MEDIAN(spqs125.time / ppq125.time) AS spqs_ratio,
    %% MEDIAN(sort125.time / ppq125.time) AS sort_ratio,
    %% MEDIAN(spq129.time / ppq129.time) AS spq_ratio,
    %% MEDIAN(spqs129.time / ppq129.time) AS spqs_ratio,
    %% MEDIAN(sort129.time / ppq129.time) AS sort_ratio
    %% FROM stats125 s
    %% LEFT JOIN oporder opo ON opo.name = s.op
    %% LEFT JOIN stats125 ppq125  ON ppq125.pqname='ppq'     AND ppq125.op = s.op  AND ppq125.num_elements = s.num_elements
    %% LEFT JOIN stats125 spq125  ON spq125.pqname='spq'     AND spq125.op = s.op  AND spq125.num_elements = s.num_elements
    %% LEFT JOIN stats125 spqs125 ON spqs125.pqname='spqs'   AND spqs125.op = s.op AND spqs125.num_elements = s.num_elements
    %% LEFT JOIN stats125 sort125 ON sort125.pqname='sorter' AND sort125.op = s.op AND sort125.num_elements = s.num_elements
    %% LEFT JOIN stats129 ppq129  ON ppq129.pqname='ppq'     AND ppq129.op = s.op  AND ppq129.num_elements = s.num_elements
    %% LEFT JOIN stats129 spq129  ON spq129.pqname='spq'     AND spq129.op = s.op  AND spq129.num_elements = s.num_elements
    %% LEFT JOIN stats129 spqs129 ON spqs129.pqname='spqs'   AND spqs129.op = s.op AND spqs129.num_elements = s.num_elements
    %% LEFT JOIN stats129 sort129 ON sort129.pqname='sorter' AND sort129.op = s.op AND sort129.num_elements = s.num_elements
    %% WHERE s.pqname='ppq' AND s.op IN ('push-rand-pop', 'push-asc-pop', 'rbulk-pop-push|cycles')
    %% AND LOG(2, s.num_elements) >= 30.5
    %% GROUP BY opo.order, opo.desc
    %% ORDER BY opo.order
        push-rand-pop & 1.39 & 3.58 & 0.87 & 2.25 & 4.83 & 0.83 \\
         push-asc-pop & 1.81 & 3.40 & 1.37 & 4.29 & 6.71 & 1.20 \\
    asc-rbulk-rewrite & 1.89 & 4.70 &      & 2.91 & 3.43 &      \\
    % END TABULAR SELECT opo.desc, MEDIAN(spq125.time / ppq125.time) AS spq_ratio...
    \hline
  \end{tabular}
  \vspace{2ex}
  \caption{Speedup of PPQ over parallelized STXXL PQ (SPQP), sequential STXXL PQ (SPQS), and STXXL Sorter for 64-bit integers, averaged for all experiments with $n \geq 2^{30.5}$.}\label{tab:speedups-8}
\end{table}

\begin{figure}[p]\centering
  \pgfplotsset{
    plotMini,
    ymin=0,
    legend to name={none},
  }
  \begin{tikzpicture}
    \begin{axis}[
      title={\bf Platform A-Rot},
      xlabel={number of items [$\log_2(n)$]},
      ylabel={\clap{million items per second}},
      ]

      \draw[EMline] (axis cs:30.5,\pgfkeysvalueof{/pgfplots/ymax}) -- (axis cs:30.5,\pgfkeysvalueof{/pgfplots/ymin});

      %% MULTIPLOT(pqname)
      %% SELECT LOG(2, num_elements) AS x, MEDIAN(num_elements / time / 1e6) AS y, MULTIPLOT
      %% FROM stats125 s LEFT JOIN pqorder p ON p.name = s.pqname
      %% WHERE op='push-rand-pop' GROUP BY p.order,MULTIPLOT,x ORDER BY p.order,MULTIPLOT,x
      \addplot coordinates { (27,26.5696) (28,24.3001) (29,22.9282) (30,22.5727) (31,17.8932) (32,16.9242) (33,17.1245) (34,17.8427) (35,17.6864) };
      \addlegendentry{pqname=ppq};
      \addplot coordinates { (27,14.5672) (28,14.3612) (29,13.2216) (30,13.1473) (31,12.7309) (32,13.0956) (33,12.4596) (34,12.8383) (35,11.7916) };
      \addlegendentry{pqname=spq};
      \addplot coordinates { (27,5.64149) (28,5.43826) (29,5.28554) (30,4.93203) (31,4.96193) (32,4.98617) (33,4.82249) (34,4.96574) (35,4.93581) };
      \addlegendentry{pqname=spqs};
      \addplot coordinates { (27,19.9943) (28,19.8976) (29,19.9408) (30,19.22) (31,20.6844) (32,21.8076) (33,19.2158) (34,15.7919) (35,20.7336) };
      \addlegendentry{pqname=sorter};

    \end{axis}
  \end{tikzpicture}
  \hfill%
  \begin{tikzpicture}
    \begin{axis}[
      title={\bf Platform B-SSD},
      xlabel={number of items [$\log_2(n)$]},
      ylabel right={\bf push-rand-pop},
      legend to name=mylegend,
      legend columns=2,
      legend transposed=true,
      ]

      \draw[EMline] (axis cs:30.5,\pgfkeysvalueof{/pgfplots/ymax}) -- (axis cs:30.5,\pgfkeysvalueof{/pgfplots/ymin});

      %% MULTIPLOT(pqname)
      %% SELECT LOG(2, num_elements) AS x, MEDIAN(num_elements / time / 1e6) AS y, MULTIPLOT
      %% FROM stats129 s LEFT JOIN pqorder p ON p.name = s.pqname
      %% WHERE op='push-rand-pop' GROUP BY p.order,MULTIPLOT,x ORDER BY p.order,MULTIPLOT,x
      \addplot coordinates { (27,30.4828) (28,25.3674) (29,18.9404) (30,19.8373) (31,28.8279) (32,33.3103) (33,28.0026) (34,25.6384) (35,25.4183) };
      \addlegendentry{pqname=ppq};
      \addplot coordinates { (27,15.1143) (28,13.7951) (29,11.6774) (30,13.7389) (31,13.1255) (32,12.9566) (33,12.5942) (34,11.5209) (35,9.97533) };
      \addlegendentry{pqname=spq};
      \addplot coordinates { (27,6.70077) (28,6.50581) (29,6.26265) (30,5.81256) (31,5.85397) (32,5.83667) (33,5.79653) (34,5.8035) (35,5.75853) };
      \addlegendentry{pqname=spqs};
      \addplot coordinates { (27,29.1869) (28,30.6969) (29,31.708) (30,31.9814) (31,32.8356) (32,33.9566) (33,33.566) (34,32.6463) (35,31.781) };
      \addlegendentry{pqname=sorter};

      \legend{Our PPQ, Parallelized STXXL PQ, Sequential STXXL PQ, STXXL Sorter}

    \end{axis}
  \end{tikzpicture}

  \begin{tikzpicture}
    \begin{axis}[
      xlabel={number of items [$\log_2(n)$]},
      ylabel={\clap{million items per second}},
      ]

      \draw[EMline] (axis cs:30.5,\pgfkeysvalueof{/pgfplots/ymax}) -- (axis cs:30.5,\pgfkeysvalueof{/pgfplots/ymin});

      %% MULTIPLOT(pqname)
      %% SELECT LOG(2, num_elements) AS x, MEDIAN(num_elements / time / 1e6) AS y, MULTIPLOT
      %% FROM stats125 s LEFT JOIN pqorder p ON p.name = s.pqname
      %% WHERE op='push-asc-pop' GROUP BY p.order,MULTIPLOT,x ORDER BY p.order,MULTIPLOT,x
      \addplot coordinates { (27,67.4763) (28,65.4408) (29,61.8639) (30,57.296) (31,23.4957) (32,23.719) (33,23.5451) (34,23.5295) (35,23.8608) };
      \addlegendentry{pqname=ppq};
      \addplot coordinates { (27,17.1412) (28,17.2003) (29,16.9642) (30,15.4892) (31,14.1265) (32,13.5664) (33,12.9902) (34,12.8099) (35,12.5964) };
      \addlegendentry{pqname=spq};
      \addplot coordinates { (27,8.32361) (28,8.29657) (29,8.2507) (30,7.42802) (31,7.26303) (32,7.13523) (33,6.91713) (34,6.8811) (35,6.81204) };
      \addlegendentry{pqname=spqs};
      \addplot coordinates { (27,18.9868) (28,18.7659) (29,18.5433) (30,15.6796) (31,16.7429) (32,17.3674) (33,17.4397) (34,17.1889) (35,16.5552) };
      \addlegendentry{pqname=sorter};

    \end{axis}
  \end{tikzpicture}
  \hfill%
  \begin{tikzpicture}
    \begin{axis}[
      xlabel={number of items [$\log_2(n)$]},
      ylabel right={\bf push-asc-pop},
      legend to name=mylegend,
      legend columns=2,
      legend transposed=true,
      ]

      \draw[EMline] (axis cs:30.5,\pgfkeysvalueof{/pgfplots/ymax}) -- (axis cs:30.5,\pgfkeysvalueof{/pgfplots/ymin});

      %% MULTIPLOT(pqname)
      %% SELECT LOG(2, num_elements) AS x, MEDIAN(num_elements / time / 1e6) AS y, MULTIPLOT
      %% FROM stats129 s LEFT JOIN pqorder p ON p.name = s.pqname
      %% WHERE op='push-asc-pop' GROUP BY p.order,MULTIPLOT,x ORDER BY p.order,MULTIPLOT,x
      \addplot coordinates { (27,106.654) (28,86.4128) (29,72.761) (30,81.7441) (31,67.6842) (32,68.3055) (33,68.1314) (34,65.4681) (35,64.746) };
      \addlegendentry{pqname=ppq};
      \addplot coordinates { (27,17.7143) (28,17.4303) (29,16.938) (30,16.5577) (31,15.9342) (32,15.762) (33,15.5095) (34,15.3694) (35,15.2089) };
      \addlegendentry{pqname=spq};
      \addplot coordinates { (27,11.5352) (28,11.4974) (29,11.403) (30,10.0786) (31,10.1274) (32,10.111) (33,10.0011) (34,9.97832) (35,9.976) };
      \addlegendentry{pqname=spqs};
      \addplot coordinates { (27,45.9352) (28,48.7443) (29,50.5257) (30,51.045) (31,54.5268) (32,58.1883) (33,57.0214) (34,54.6928) (35,52.7736) };
      \addlegendentry{pqname=sorter};

      \legend{Our PPQ, Parallelized STXXL PQ, Sequential STXXL PQ, STXXL Sorter}

    \end{axis}
  \end{tikzpicture}

  \begin{tikzpicture}
    \begin{axis}[
      xlabel={number of items [$\log_2(n)$]},
      ylabel={\clap{million items per second}},
      ]

      \draw[EMline] (axis cs:30.5,\pgfkeysvalueof{/pgfplots/ymax}) -- (axis cs:30.5,\pgfkeysvalueof{/pgfplots/ymin});

      %% MULTIPLOT(pqname)
      %% SELECT LOG(2, num_elements) AS x, MEDIAN(num_elements / time / 1e6) AS y, MULTIPLOT
      %% FROM stats125 WHERE op='rbulk-pop-push|cycles' GROUP BY MULTIPLOT,x ORDER BY MULTIPLOT,x
      \addplot coordinates { (27,134.033) (28,113.044) (29,99.414) (30,110.62) (31,172.195) (32,69.2787) (33,69.7067) (34,69.2584) (35,65.5519) };
      \addlegendentry{pqname=ppq};
      \addplot coordinates { (27,50.1938) (28,48.3974) (29,47.0367) (30,44.8805) (31,37.4941) (32,37.318) (33,37.0346) (34,36.0057) (35,34.9587) };
      \addlegendentry{pqname=spq};
      \addplot coordinates { (27,16.4215) (28,16.2549) (29,16.1545) (30,16.484) (31,15.284) (32,15.0397) (33,14.7845) (34,14.6003) (35,14.3895) };
      \addlegendentry{pqname=spqs};

    \end{axis}
  \end{tikzpicture}
  \hfill%
  \begin{tikzpicture}
    \begin{axis}[
      xlabel={number of items [$\log_2(n)$]},
      ylabel right={\bf asc-rbulk-rewrite},
      ]

      \draw[EMline] (axis cs:30.5,\pgfkeysvalueof{/pgfplots/ymax}) -- (axis cs:30.5,\pgfkeysvalueof{/pgfplots/ymin});

      %% MULTIPLOT(pqname)
      %% SELECT LOG(2, num_elements) AS x, MEDIAN(num_elements / time / 1e6) AS y, MULTIPLOT
      %% FROM stats129 WHERE op='rbulk-pop-push|cycles' GROUP BY MULTIPLOT,x ORDER BY MULTIPLOT,x
      \addplot coordinates { (27,88.4373) (28,75.3213) (29,86.6214) (30,85.9702) (31,185.732) (32,173.679) (33,147.18) (34,159.535) (35,159.093) };
      \addlegendentry{pqname=ppq};
      \addplot coordinates { (27,68.867) (28,65.4454) (29,58.4098) (30,81.5059) (31,63.9764) (32,61.4334) (33,58.3797) (34,54.3124) (35,48.2925) };
      \addlegendentry{pqname=spq};
      \addplot coordinates { (27,63.8825) (28,61.9902) (29,59.1717) (30,59.6344) (31,51.4056) (32,49.3043) (33,49.3045) (34,46.6497) (35,46.643) };
      \addlegendentry{pqname=spqs};

    \end{axis}
  \end{tikzpicture}

  \begin{tikzpicture}
    \begin{axis}[
      plotMiniBulkSize,
      xlabel={items in bulk sequences [$\log_2(v)$]},
      ylabel={\clap{million items per second}},
      ]

      %% MULTIPLOT(pqname)
      %% SELECT LOG(2, bulk_size) AS x, MEDIAN(num_elements / time / 1e6) AS y, MULTIPLOT
      %% FROM stats125b s LEFT JOIN pqorder p ON p.name = s.pqname
      %% WHERE op='bulk-pop-push|cycles' GROUP BY p.order,MULTIPLOT,x ORDER BY p.order,MULTIPLOT,x
      \addplot coordinates { (12.2877,18.9306) (13.2877,29.6719) (14.2877,42.2226) (15.2877,51.9967) (16.2877,59.6347) (17.2877,64.9549) (18.2877,69.725) (19.2877,68.1964) (20.2877,67.1715) (21.2877,67.1072) (22.2877,65.8141) };
      \addlegendentry{pqname=ppq};
      \addplot coordinates { (12.2877,37.743) (13.2877,37.7963) (14.2877,37.7044) (15.2877,37.6826) (16.2877,37.6707) (17.2877,37.4796) (18.2877,37.3185) (19.2877,36.5755) (20.2877,36.287) (21.2877,32.3867) (22.2877,31.8312) };
      \addlegendentry{pqname=spq};
      \addplot coordinates { (12.2877,15.0332) (13.2877,15.0091) (14.2877,15.0898) (15.2877,15.0783) (16.2877,15.0515) (17.2877,15.0297) (18.2877,15.0395) (19.2877,15.0603) (20.2877,14.3879) (21.2877,13.58) (22.2877,13.5505) };
      \addlegendentry{pqname=spqs};

    \end{axis}
  \end{tikzpicture}
  \hfill%
  \begin{tikzpicture}
    \begin{axis}[
      plotMiniBulkSize,
      xlabel={items in bulk sequences [$\log_2(v)$]},
      ylabel right={\bf bulk-rewrite},
      ]

      %% MULTIPLOT(pqname)
      %% SELECT LOG(2, bulk_size) AS x, MEDIAN(num_elements / time / 1e6) AS y, MULTIPLOT
      %% FROM stats129b s LEFT JOIN pqorder p ON p.name = s.pqname
      %% WHERE op='bulk-pop-push|cycles' GROUP BY p.order,MULTIPLOT,x ORDER BY p.order,MULTIPLOT,x
      \addplot coordinates { (12.2877,10.2678) (13.2877,18.3502) (14.2877,32.4112) (15.2877,57.0125) (16.2877,92.9449) (17.2877,138.071) (18.2877,188.933) (19.2877,230.721) (20.2877,255.813) (21.2877,273.439) (22.2877,229.334) };
      \addlegendentry{pqname=ppq};
      \addplot coordinates { (12.2877,61.2427) (13.2877,61.5581) (14.2877,60.8852) (15.2877,60.4222) (16.2877,62.9854) (17.2877,61.4834) (18.2877,62.1683) (19.2877,63.9037) (20.2877,59.5779) (21.2877,58.5208) (22.2877,59.5158) };
      \addlegendentry{pqname=spq};
      \addplot coordinates { (12.2877,51.5454) (13.2877,51.964) (14.2877,52.2925) (15.2877,52.255) (16.2877,52.2177) (17.2877,51.0697) (18.2877,52.3197) (19.2877,47.9655) (20.2877,48.2584) (21.2877,41.7786) (22.2877,42.5143) };
      \addlegendentry{pqname=spqs};

    \end{axis}
  \end{tikzpicture}

  \centerline{\ref{mylegend}}

  \caption{Experimental results of our four benchmarks with 64-bit integer items.}\label{fig:results-8}
\end{figure}
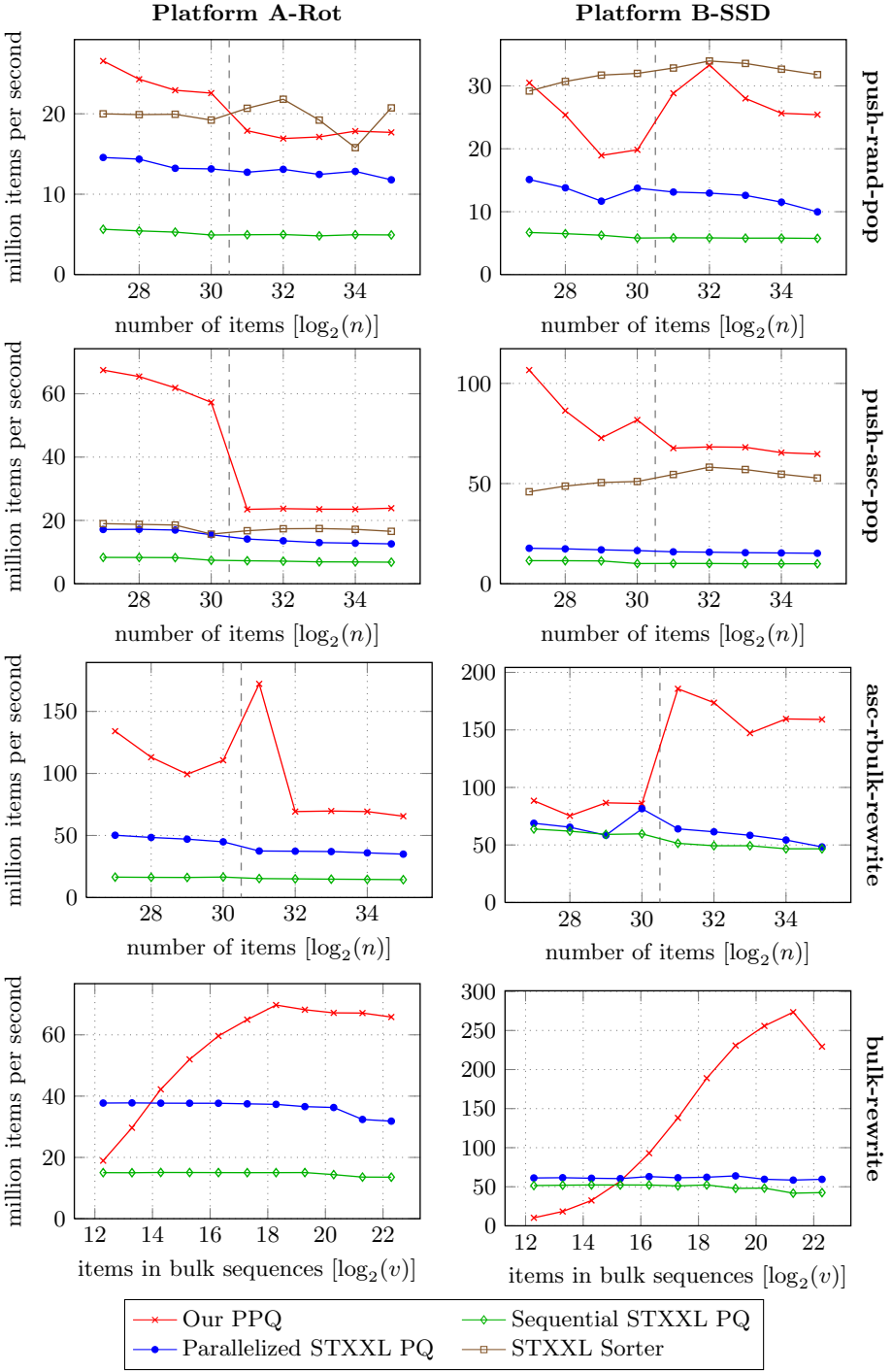

In all our experiments, except the bulk size benchmark, our PPQ is faster than
the parallelized and sequential STXXL PQ. Assuming the PQs use
12\,GiB of the 16\,GiB RAM for storing items, then the containers only need EM
for about $n \geq 2^{30.5}$ (indicated by dashed horizontal line in plots). In
Table~\ref{tab:speedups-8} we show the average execution time speedups of our PPQ
for the available competitors, averaged over all inputs where the input cannot
fit into RAM. Remarkably, on both platforms the PPQ is faster than the sorter
for both inputs except random on A-Rot, which indicates that I/O overlaps
computation work very well, often even better than the sorter. We may
investigate how to increase the sorter's speed using our techniques. Comparing
to SPQS, we achieved speedups of 3.6\,--\,4.7 on A-Rot (which has 8 real cores),
and speedups of 3.4--6.7 on B-SSD (16 real cores). Compared to the previously
parallelized SPQP, we only gain 1.4\,--1.9 on A-Rot and 2.2\,--\,4.3 speedup on
B-SSD. While this relative comparison may not seem much, by comparing the PPQ's
throughput to the sorter and the absolute I/O bandwidth of the disks
(Table~\ref{tab:results-8} in appendix), one can see that the PPQ reaches 64\%
of the available I/O bandwidth in \emph{push-asc-pop} on B-SSD, and 49\% on A-Rot.
For \emph{asc-rbulk-rewrite} the PQ-throughput is naturally higher than the possible I/O bandwidth,
since the PQs keep items in RAM. In \emph{push-rand-pop}, the PPQ is
limited by compute time of sorting random integers, just as the STXXL sorter
is. For \emph{asc-rbulk-rewrite}, which is a main focus of the PPQ, we achieve a
speedup of 1.9 on A-Rot and 2.7 on B-SSD for bulk sizes of on average 320\,000 items.
Considering the increasing bulk sizes in the \emph{bulk-rewrite} experiment, we
see that larger bulks yield better performance up to a certain sweet spot (on
B-SSD it is probably even higher), but the break even of the PPQ over the SPQP
is quite low: $20\,\text{K}$ items for A-Rot and $80\,\text{K}$ items for B-SSD.

% ------------------------------------------------------------------------------
\section{Conclusions and Future Work}\label{sec:conclusions}

We presented a PPQ design and implementation for EM, and successfully
demonstrated that for specific benchmarks the high I/O bandwidth of parallel
disks and even SSDs can be utilized. By relaxing semantics, our bulk-parallel
interface enables parallelized processing of larger amounts of items in the PPQ.
In the future, we want to apply our PPQ's bulk-parallel processing to the eSAIS
external suffix and LCP sorting algorithm~\cite{bingmann2013inducing}, where in
the largest recursion level each alphabet character (and repetition count) is a
bulk.

During our work on the PPQ two important issues remained untouched: how does one balance
work in an EM algorithm library when the user application, the EM containers,
and I/O overlapping require parallel work? We left this to the operating system
scheduler and block the user application during parallel merging, which is not
desirable. As indicated by theory and experiments, \emph{bulk\_pop\_limit}
requires large bulks to work efficiently, however, the PPQ cannot know the
resulting bulk sizes without performing a costly multisequence selection. One
could require the user application to provide an estimate of the resulting size, or
develop an online oracle.  Finally, experiments with other internal memory
PPQs and $d$-ary heaps may improve performance by using larger insertion heaps.

\bibliographystyle{splncs03}
\bibliography{library.bib}

% ------------------------------------------------------------------------------
\clearpage\appendix
\renewcommand{\textfraction}{0.05}
\renewcommand{\topfraction}{0.8}
\renewcommand{\bottomfraction}{0.8}

\section{Performance of Parallel Multiway-Merge}\label{sec:multiwaymerge}

Folk cache-efficiency wisdom dictates that one should only merge a small number
of sequences using tournament trees, such that the data structures and heads of
sequences fit into cache. The EM sequence heap implementation in the STXXL
merges a fixed maximum of 64 sequences.

\begin{figure}[b!]\centering
  \tikzset{
    line cap=round,
  }%
  \pgfplotsset{
    width=63mm,height=44mm,
    cycle list={
      red, every mark/.append style={solid,scale=0.6}, mark=x \\%
      blue, every mark/.append style={solid,scale=0.4}, mark=* \\%
    },
    ymin=0,
    title style={yshift=-3pt},
  }%
  \vspace{-2em}
  \begin{tikzpicture}
    \begin{semilogxaxis}[
      title={64-bit integers},
      xlabel={number of 2\,MiB sequences [$v$]},
      ylabel={\clap{time per item [ns/item]}},
      axis y line*=left,
      ]

      %% PLOT SELECT seq_count AS x, MEDIAN(time / total_size / inner_repeats * 1e9) AS y
      %% FROM paramwm1 WHERE item_size = 8 GROUP BY x ORDER BY x
      \addplot coordinates { (1,1.11618) (2,2.44425) (3,0.950882) (4,1.18407) (5,5.09379) (6,5.45001) (7,5.93649) (8,5.90485) (9,6.24366) (10,4.80671) (11,5.18544) (12,4.62292) (13,4.80718) (14,4.82362) (15,3.78448) (16,4.45809) (17,4.43848) (18,4.44315) (19,4.43327) (20,4.56591) (21,4.81875) (22,4.47979) (23,4.79004) (24,4.2909) (25,4.36785) (26,4.22823) (27,4.44167) (28,4.56823) (29,4.46178) (30,4.49075) (31,4.48348) (32,4.44057) (33,4.58045) (34,4.5343) (35,4.64986) (36,4.68985) (37,4.74193) (38,4.93816) (39,4.96224) (40,4.67389) (41,4.73046) (42,4.76834) (43,4.75334) (44,4.86603) (45,4.79335) (46,4.83834) (47,4.80853) (48,4.85845) (49,4.73968) (50,4.77207) (51,4.8042) (52,4.84596) (53,4.87717) (54,4.94678) (55,4.98286) (56,4.92019) (57,4.79883) (58,4.83844) (59,4.88356) (60,4.86481) (61,4.85959) (62,4.89659) (63,4.91878) (64,4.84709) (66,5.09594) (68,5.08698) (70,5.25383) (72,5.23461) (74,5.30865) (76,5.39642) (78,5.32181) (80,5.33795) (82,5.32383) (84,5.40512) (86,5.44409) (88,5.4975) (90,5.46612) (92,5.5042) (94,5.49362) (96,5.58088) (98,5.53958) (100,5.48077) (102,5.55899) (104,5.55226) (106,5.62402) (108,5.68193) (110,5.71036) (112,5.75019) (114,5.63314) (116,5.69917) (118,5.74377) (120,5.72161) (122,5.63395) (124,5.63225) (126,5.63601) (128,5.64009) (132,6.05764) (136,6.21605) (140,6.25013) (144,6.23624) (148,6.3187) (152,6.34876) (156,6.36494) (160,6.35828) (164,6.42082) (168,6.50971) (172,6.51921) (176,6.51066) (180,6.46346) (184,6.44694) (188,6.47456) (192,6.4392) (196,6.30738) (200,6.32753) (204,6.33974) (208,6.37915) (212,6.36782) (216,6.55877) (220,6.52525) (224,6.54582) (228,6.42997) (232,6.45078) (236,6.46296) (240,6.46341) (244,6.55272) (248,6.50689) (252,6.54095) (256,6.49911) (272,6.90426) (288,6.97237) (304,7.19038) (320,7.15245) (336,7.21748) (352,7.26994) (368,7.27941) (384,7.27009) (400,7.19702) (416,7.22741) (432,7.30339) (448,7.3219) (464,7.29278) (480,7.33922) (496,7.32257) (512,7.3507) (576,7.86657) (640,7.97275) (704,8.02864) (768,8.08084) (832,8.08612) (896,8.17379) (960,8.24735) (1024,8.28404) (1152,8.90086) (1280,9.11098) (1408,9.34575) (1536,9.49387) (1664,9.60149) (1792,9.84505) (1920,10.0289) (2048,10.2183) (2176,10.7728) (2304,11.0264) (2432,11.2578) (2560,11.4091) (2688,11.6725) (2816,11.8688) (2944,12.072) (3072,12.3261) (3200,12.3427) (3328,12.4572) (3456,12.58) (3584,12.7088) (3712,12.8517) (3840,13.0487) (3968,13.1058) };

    \end{semilogxaxis}
    \begin{semilogxaxis}[
      axis y line*=right,
      axis x line=none,
      cycle list shift=1,
      ymax=6.35,
      ]

      %% PLOT SELECT seq_count AS x, MEDIAN(time / total_size / inner_repeats * 1e9) AS y
      %% FROM paramwm2 WHERE item_size = 8 GROUP BY x ORDER BY x
      \addplot coordinates { (1,1.92356) (2,1.47188) (3,0.696654) (4,0.861615) (5,4.0884) (6,3.86715) (7,3.69808) (8,3.29679) (9,3.2129) (10,3.5219) (11,2.89869) (12,2.93698) (13,3.06615) (14,3.16678) (15,3.0956) (16,2.91505) (17,2.88586) (18,3.06989) (19,2.99055) (20,3.03113) (21,3.03341) (22,3.13302) (23,2.8082) (24,3.05821) (25,3.04627) (26,3.08018) (27,2.86224) (28,3.18614) (29,3.15935) (30,3.09928) (31,3.25181) (32,3.11377) (33,3.15039) (34,3.11888) (35,3.23919) (36,3.1018) (37,2.9988) (38,3.07226) (39,3.18298) (40,3.23123) (41,3.15676) (42,3.17461) (43,3.35747) (44,3.15598) (45,3.23164) (46,3.26391) (47,3.36478) (48,3.33493) (49,3.3286) (50,3.38107) (51,3.41836) (52,3.20618) (53,3.2111) (54,3.30931) (55,3.3834) (56,3.34468) (57,3.30425) (58,3.22279) (59,3.33639) (60,3.19057) (61,3.22092) (62,3.18832) (63,3.27089) (64,3.04748) (66,3.15451) (68,3.2016) (70,3.22474) (72,3.10348) (74,3.1416) (76,3.10408) (78,3.12675) (80,3.10253) (82,3.10651) (84,3.08846) (86,3.06985) (88,3.08218) (90,3.08777) (92,3.1995) (94,3.18654) (96,3.18988) (98,3.17308) (100,3.20454) (102,3.16992) (104,3.22946) (106,3.23092) (108,3.2318) (110,3.28573) (112,3.28207) (114,3.2499) (116,3.24932) (118,3.28174) (120,3.26997) (122,3.26284) (124,3.27302) (126,3.28129) (128,3.25372) (132,3.3435) (136,3.34978) (140,3.35265) (144,3.33709) (148,3.36199) (152,3.40768) (156,3.38066) (160,3.37622) (164,3.36393) (168,3.3868) (172,3.36876) (176,3.40039) (180,3.37159) (184,3.38891) (188,3.43835) (192,3.4177) (196,3.392) (200,3.39957) (204,3.40704) (208,3.38587) (212,3.41513) (216,3.43961) (220,3.44602) (224,3.45778) (228,3.44911) (232,3.4555) (236,3.46532) (240,3.45835) (244,3.50729) (248,3.54661) (252,3.53741) (256,3.57505) (272,3.78229) (288,3.78865) (304,3.82832) (320,3.83827) (336,3.88463) (352,3.89437) (368,3.92174) (384,3.95557) (400,3.9503) (416,3.96915) (432,3.98068) (448,4.0284) (464,4.02185) (480,4.04411) (496,4.05765) (512,4.08859) (576,4.31759) (640,4.40424) (704,4.70235) (768,4.75952) (832,4.82495) (896,4.82791) (960,4.92346) (1024,4.95026) (1152,5.1045) (1280,5.14307) (1408,5.23954) (1536,5.30756) (1664,5.33266) (1792,5.29984) (1920,5.28414) (2048,5.2751) (2176,5.2756) (2304,5.26525) (2432,5.25325) (2560,5.23905) (2688,5.24067) (2816,5.31133) (2944,5.28654) (3072,5.31857) (3200,5.47378) (3328,5.4228) (3456,5.43807) (3584,5.47222) (3712,5.4903) (3840,5.50822) (3968,5.53116) };

    \end{semilogxaxis}
  \end{tikzpicture}
  \hfill%
  \begin{tikzpicture}
    \begin{semilogxaxis}[
      title={36-byte structures},
      xlabel={number of 2\,MiB sequences [$v$]},
      axis y line*=left,
      ]

      %% PLOT SELECT seq_count AS x, MEDIAN(time / total_size / inner_repeats * 1e9) AS y
      %% FROM paramwm1 WHERE item_size = 36 GROUP BY x ORDER BY x
      \addplot coordinates { (1,4.12446) (2,8.76966) (3,2.36059) (4,4.31914) (5,8.07227) (6,9.10152) (7,10.3818) (8,10.5122) (9,10.8509) (10,9.88047) (11,10.3582) (12,9.75753) (13,9.67904) (14,9.7156) (15,8.57697) (16,9.33377) (17,9.35236) (18,9.10357) (19,9.34733) (20,9.4817) (21,10.0241) (22,10.2718) (23,8.72483) (24,8.89192) (25,9.16607) (26,8.6728) (27,8.84259) (28,8.87545) (29,8.7224) (30,8.61035) (31,8.39578) (32,8.69402) (33,8.75212) (34,8.8204) (35,8.6727) (36,9.02079) (37,8.99218) (38,8.81384) (39,8.55439) (40,8.64804) (41,8.71855) (42,9.07353) (43,8.792) (44,8.88055) (45,8.99524) (46,9.06415) (47,8.83726) (48,8.8568) (49,9.07299) (50,9.06397) (51,8.95172) (52,8.98917) (53,9.06481) (54,9.08438) (55,9.02284) (56,9.00167) (57,8.8804) (58,8.8482) (59,8.85596) (60,8.87408) (61,9.24165) (62,9.20925) (63,9.08336) (64,9.1315) (66,9.2981) (68,9.25844) (70,10.0268) (72,10.1707) (74,10.3735) (76,10.4343) (78,10.2895) (80,10.3502) (82,10.1562) (84,10.3208) (86,10.2258) (88,10.2248) (90,10.252) (92,10.4464) (94,10.4272) (96,11.1116) (98,11.5167) (100,11.3584) (102,11.7255) (104,11.5129) (106,11.6544) (108,11.4863) (110,11.8956) (112,12.0753) (114,12.1024) (116,12.3484) (118,11.914) (120,11.948) (122,11.8025) (124,11.6194) (126,11.6103) (128,11.5965) (132,11.959) (136,13.3578) (140,13.2685) (144,13.6015) (148,13.6615) (152,13.5828) (156,13.8209) (160,14.0868) (164,14.1624) (168,14.4551) (172,14.3365) (176,14.0664) (180,13.7714) (184,13.83) (188,13.6963) (192,13.8383) (196,13.5146) (200,13.4141) (204,13.4012) (208,13.5518) (212,13.5696) (216,15.1365) (220,14.9612) (224,14.7211) (228,14.4586) (232,14.6364) (236,14.4465) (240,14.5357) (244,15.735) (248,15.7417) (252,15.7816) (256,15.7202) (272,15.8639) (288,15.6996) (304,17.5058) (320,17.6278) (336,17.8362) (352,18.0431) (368,18.2175) (384,18.2864) (400,18.5577) (416,18.7881) (432,18.882) (448,18.8807) (464,19.1359) (480,19.2939) (496,19.4587) (512,19.6325) (576,21.039) (640,20.1946) (704,21.857) (768,22.4193) (832,23.0221) (896,23.5373) (960,24.0399) (1024,24.5574) (1152,26.4763) (1280,27.4702) (1408,28.3852) (1536,29.508) (1664,30.8458) (1792,31.9995) (1920,33.2794) (2048,34.7254) (2176,37.2116) (2304,38.4082) (2432,39.4847) (2560,40.6059) (2688,41.8805) (2816,42.8599) (2944,43.8619) (3072,44.9003) (3200,46.0718) (3328,46.8727) (3456,47.7056) (3584,48.4284) (3712,49.4829) (3840,50.1229) (3968,50.8429) }; \label{plot_one}

    \end{semilogxaxis}
    \begin{semilogxaxis}[
      axis y line*=right,
      axis x line=none,
      cycle list shift=1,
      legend pos=north west,
      ]

      \addlegendimage{/pgfplots/refstyle=plot_one}

      %% PLOT SELECT seq_count AS x, MEDIAN(time / total_size / inner_repeats * 1e9) AS y
      %% FROM paramwm2 WHERE item_size = 36 GROUP BY x ORDER BY x
      \addplot coordinates { (1,3.94772) (2,10.9388) (3,2.33679) (4,2.34616) (5,6.58712) (6,6.2179) (7,6.38091) (8,5.67879) (9,5.8346) (10,5.58406) (11,5.26029) (12,5.28614) (13,5.34519) (14,5.52246) (15,5.58747) (16,5.07172) (17,5.15156) (18,5.12086) (19,5.46584) (20,5.64537) (21,5.75949) (22,5.90063) (23,6.12535) (24,6.01019) (25,5.96787) (26,5.9545) (27,5.88229) (28,5.83653) (29,5.79308) (30,5.77999) (31,6.01095) (32,6.15268) (33,6.21724) (34,6.31798) (35,6.27463) (36,6.12256) (37,6.18585) (38,6.17464) (39,6.19778) (40,5.94614) (41,6.0252) (42,6.17483) (43,5.9985) (44,6.09339) (45,5.98677) (46,5.9298) (47,6.02655) (48,6.10378) (49,6.20069) (50,6.11036) (51,6.33156) (52,5.99399) (53,5.85842) (54,5.8677) (55,6.02465) (56,6.00769) (57,5.87444) (58,6.00891) (59,6.11223) (60,6.01926) (61,5.95906) (62,6.18615) (63,6.31599) (64,6.09456) (66,6.30956) (68,6.58481) (70,6.89246) (72,7.12157) (74,7.12166) (76,7.13427) (78,7.26916) (80,7.18842) (82,7.27205) (84,7.25223) (86,7.11585) (88,7.03005) (90,6.90995) (92,6.77545) (94,6.65106) (96,7.17353) (98,7.51126) (100,7.6975) (102,7.59426) (104,7.92941) (106,7.9625) (108,7.90107) (110,8.19305) (112,8.32559) (114,8.12072) (116,8.17998) (118,8.17903) (120,8.1771) (122,8.15773) (124,8.13313) (126,8.12187) (128,8.27486) (132,8.39372) (136,8.40884) (140,8.32362) (144,8.23501) (148,8.29484) (152,8.06497) (156,8.18163) (160,8.19971) (164,8.29512) (168,8.1399) (172,8.16708) (176,8.09143) (180,8.00246) (184,8.08371) (188,8.05636) (192,7.88508) (196,7.76596) (200,7.77296) (204,7.69132) (208,7.54931) (212,7.77742) (216,7.50552) (220,7.71316) (224,7.48674) (228,7.7634) (232,7.64893) (236,7.77241) (240,7.58507) (244,8.18657) (248,7.9332) (252,8.15744) (256,7.98689) (272,8.24625) (288,8.3434) (304,8.47285) (320,8.60117) (336,8.59971) (352,8.65093) (368,8.6602) (384,8.85614) (400,8.8497) (416,9.03711) (432,9.19814) (448,9.31592) (464,9.4667) (480,9.61285) (496,9.64249) (512,9.78219) (576,10.4744) (640,10.894) (704,14.6944) (768,14.9625) (832,15.0515) (896,15.4316) (960,15.813) (1024,16.0596) (1152,16.7658) (1280,17.5152) (1408,17.7193) (1536,17.8861) (1664,18.5563) (1792,18.8077) (1920,18.9521) (2048,18.7188) (2176,20.3721) (2304,19.739) (2432,21.013) (2560,20.7373) (2688,14.3278) (2816,14.3724) (2944,14.9154) (3072,14.6575) (3200,14.9673) (3328,14.9787) (3456,15.2338) (3584,15.2437) (3712,15.4583) (3840,15.6034) (3968,16.009) };

      \legend{A-Rot,B-SSD}

    \end{semilogxaxis}
  \end{tikzpicture}
  \vspace{-2ex}
  \caption{Speed of parallel multiway merging in RAM, median of at least 15 repetitions. The left y axis ticks are for platform A-Rot, the right ones for platform B-SSD.}\label{fig:pmwm-results}
\end{figure}
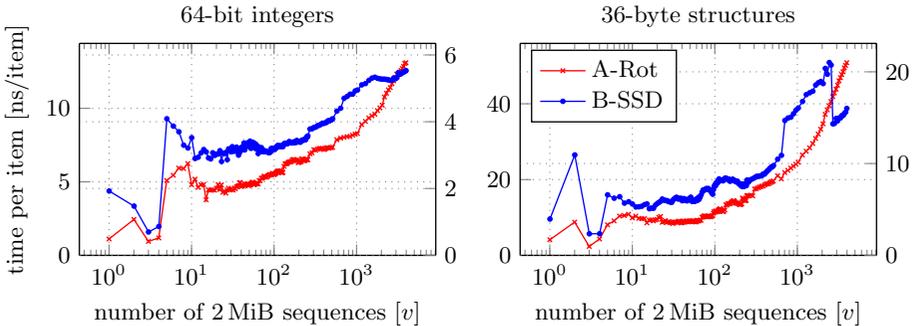

For parallel multiway merging \cite{singler2007mcstl} we have to
reevaluate these assumptions to find good balances. Due to the asymptotic
complexities discussed at the end of Section~\ref{sec:related}, it is obvious
that merging longer sequences is better, since the work is divided evenly among
processors and only effects the sequential runtime logarithmically. However, due
to cache effects and critical sequential path, it is unclear how the number of
sequences effects performance, so we performed a basic experiment merging an
increasing number $v$ of fixed length sequences of 2\,MiB size in RAM on our
platforms. The results for 64-bit integers and a larger structure are shown in
Figure~\ref{fig:pmwm-results}.

Obviously the merging speed decreases with the number of sequences, but we
cannot determine any precise number of sequences that is best: the range 10--100
seems acceptable for both data types. It is not obvious how fast merging larger
number of sequences in two steps would be, since in the second step longer
sequences are merged.

% IMPORT-DATA paramwm1 plotdata/paramwm/i10pc125/stats*.txt
% IMPORT-DATA paramwm2 plotdata/paramwm/i10pc129/stats*.txt

\begin{table}[b!]
  \begin{tabularx}{\linewidth}{|l|X|r|r|r|r|r|r|}
    \hline
    name  & processor,                     & clock  & sockets $\times$      & cache: L1     & L2             & L3    & RAM   \\
          & host bus adapter               & [GHz]  & cores $\times$ HT     & [KiB]         & [KiB]          & [MiB] & [GiB] \\
    \hline
    A-Rot & Intel Xenon X5550,             & 2.66   & $2 \times 4 \times 2$ & $4 \times 64$ & $4 \times 256$ & $8$   & 48    \\
%         & Adaptec ASR-5805               &        &                       &               &                &       &       \\
    \hline
    B-SSD & Intel Xenon E5-2650 v2,        & 2.6    & $2 \times 8 \times 2$ & $8 \times 64$ & $8 \times 256$ & $20$  & 128   \\
%         & Adaptec ASA-7805H              &        &                       &               &                &       &       \\
    \hline
  \end{tabularx}
  \vspace{0.5cm}
  \begin{tabularx}{\linewidth}{|l|X|l|r|r|r|}
    \hline
    name  & hard drives                    & link   & capacity              & R/W speed     & parallel speed                 \\
          &                                &        & [GB]                  & [MiB/s]       & [MiB/s]                        \\
    \hline
    A-Rot & 8 $\times$ WD Blue WD10EZEX    & SATAv2 & 8 $\times$ 1000       & 170\,--\,85   & 748\,--\,731                   \\
    \hline
    B-SSD & 4 $\times$ Samsung SSD 840 EVO & SATAv3 & 4 $\times$ 1000       & 512\,R 475\,W & 2\,006\,R 1\,616\,W            \\
    \hline
  \end{tabularx}
  \vspace{-2ex}
  \caption{Hardware characteristics of experimental platforms. Read/write speeds vary on rotational disks due to geometry, and on SSDs due to circuitry.}\label{tbl:platforms}
\end{table}

\begin{table}\centering
  \begin{tabular}{|r|r|*4{r}|*4{r}|*3{r}|}
    \hline
    items      & item vol. & \multicolumn{4}{c|}{push-rand-pop} & \multicolumn{4}{c|}{push-asc-pop} & \multicolumn{3}{c|}{asc-rbulk-rewrite}                     \\
    $\log_2 n$ & [GiB]     & PPQ                                & SPQP                              & SPQS & Sort & PPQ & SPQP & SPQS & Sort & PPQ & SPQP & SPQS \\ \hline
    \multicolumn{2}{|c|}{}          & \multicolumn{11}{c|}{Platform A-Rot}                                                                                                \\
    \hline
    27 &   1 & 405 & 222 & 86 & 305 & 1\,030 & 262 & 127 & 290 & 2\,045 & 766 & 251 \\
    28 &   2 & 371 & 219 & 83 & 304 &    999 & 262 & 127 & 286 & 1\,725 & 738 & 248 \\
    29 &   4 & 350 & 202 & 81 & 304 &    944 & 259 & 126 & 283 & 1\,517 & 718 & 246 \\
    30 &   8 & 344 & 201 & 75 & 293 &    874 & 236 & 113 & 239 & 1\,688 & 685 & 252 \\ \hdashline
    31 &  16 & 273 & 194 & 76 & 316 &    359 & 216 & 111 & 255 & 2\,627 & 572 & 233 \\
    32 &  32 & 258 & 200 & 76 & 333 &    362 & 207 & 109 & 265 & 1\,057 & 569 & 229 \\
    33 &  64 & 261 & 190 & 74 & 293 &    359 & 198 & 106 & 266 & 1\,064 & 565 & 226 \\
    34 & 128 & 272 & 196 & 76 & 241 &    359 & 195 & 105 & 262 & 1\,057 & 549 & 223 \\
    35 & 256 & 270 & 180 & 75 & 316 &    364 & 192 & 104 & 253 & 1\,000 & 533 & 220 \\
    % END TABULAR SELECT LOG(2, num_elements) AS logsize, num_elements * value_si...
    \hline
    \multicolumn{2}{|c|}{} & \multicolumn{11}{c|}{Platform B-SSD}                                       \\
    \hline
    27 &   1 & 465 & 231 & 102 & 445 & 1\,627 & 270 & 176 & 701 & 1\,349 & 1\,051 & 975 \\
    28 &   2 & 387 & 210 &  99 & 468 & 1\,319 & 266 & 175 & 744 & 1\,149 &    999 & 946 \\
    29 &   4 & 289 & 178 &  96 & 484 & 1\,110 & 258 & 174 & 771 & 1\,322 &    891 & 903 \\
    30 &   8 & 303 & 210 &  89 & 488 & 1\,247 & 253 & 154 & 779 & 1\,312 & 1\,244 & 910 \\ \hdashline
    31 &  16 & 440 & 200 &  89 & 501 & 1\,033 & 243 & 155 & 832 & 2\,834 &    976 & 784 \\
    32 &  32 & 508 & 198 &  89 & 518 & 1\,042 & 241 & 154 & 888 & 2\,650 &    937 & 752 \\
    33 &  64 & 427 & 192 &  88 & 512 & 1\,040 & 237 & 153 & 870 & 2\,246 &    891 & 752 \\
    34 & 128 & 391 & 176 &  89 & 498 &    999 & 235 & 152 & 835 & 2\,434 &    829 & 712 \\
    35 & 256 & 388 & 152 &  88 & 485 &    988 & 232 & 152 & 805 & 2\,428 &    737 & 712 \\
    % END TABULAR SELECT LOG(2, num_elements), num_elements * value_size / 1024 /...
    \hline
  \end{tabular}
  \medskip

  \def\tabcolsep{6pt}
  \begin{tabular}{|r|r|*3{r}|*3{r}|}
    \hline
    bulk size  & bulk vol. & \multicolumn{3}{c|}{A-Rot} & \multicolumn{3}{c|}{B-SSD}      \\
    $n / 1000$ & [KiB]     & PPQ                        & SPQP & SPQS & PPQ & SPQP & SPQS \\
    \hline
    %% TABULAR REFORMAT(precision=0 group=\,)
    %% SELECT bulk_size / 1000, bulk_size * value_size / 1024,
    %% (SELECT MEDIAN( num_elements * value_size * 2 / time / 1024 / 1024 ) FROM stats125b b WHERE b.pqname='ppq' AND op='bulk-pop-push|cycles' AND b.bulk_size = a.bulk_size),
    %% (SELECT MEDIAN( num_elements * value_size * 2 / time / 1024 / 1024 ) FROM stats125b b WHERE b.pqname='spq' AND op='bulk-pop-push|cycles' AND b.bulk_size = a.bulk_size),
    %% (SELECT MEDIAN( num_elements * value_size * 2 / time / 1024 / 1024 ) FROM stats125b b WHERE b.pqname='spqs' AND op='bulk-pop-push|cycles' AND b.bulk_size = a.bulk_size),
    %% (SELECT MEDIAN( num_elements * value_size * 2 / time / 1024 / 1024 ) FROM stats129b b WHERE b.pqname='ppq' AND op='bulk-pop-push|cycles' AND b.bulk_size = a.bulk_size),
    %% (SELECT MEDIAN( num_elements * value_size * 2 / time / 1024 / 1024 ) FROM stats129b b WHERE b.pqname='spq' AND op='bulk-pop-push|cycles' AND b.bulk_size = a.bulk_size),
    %% (SELECT MEDIAN( num_elements * value_size * 2 / time / 1024 / 1024 ) FROM stats129b b WHERE b.pqname='spqs' AND op='bulk-pop-push|cycles' AND b.bulk_size = a.bulk_size)
    %% FROM stats125b a GROUP BY bulk_size, num_elements, value_size ORDER BY bulk_size
         5 &      39 &    289 & 576 & 229 &    157 & 934 & 787 \\
        10 &      78 &    453 & 577 & 229 &    280 & 939 & 793 \\
        20 &     156 &    644 & 575 & 230 &    495 & 929 & 798 \\
        40 &     312 &    793 & 575 & 230 &    870 & 922 & 797 \\
        80 &     625 &    910 & 575 & 230 & 1\,418 & 961 & 797 \\
       160 &  1\,250 &    991 & 572 & 229 & 2\,107 & 938 & 779 \\
       320 &  2\,500 & 1\,064 & 569 & 229 & 2\,883 & 949 & 798 \\
       640 &  5\,000 & 1\,041 & 558 & 230 & 3\,521 & 975 & 732 \\
    1\,280 & 10\,000 & 1\,025 & 554 & 220 & 3\,903 & 909 & 736 \\
    2\,560 & 20\,000 & 1\,024 & 494 & 207 & 4\,172 & 893 & 637 \\
    5\,120 & 40\,000 & 1\,004 & 486 & 207 & 3\,499 & 908 & 649 \\
    % END TABULAR SELECT bulk_size / 1000, bulk_size * value_size / 1024, (SELECT...
    \hline
  \end{tabular}
  \medskip
  \caption{Experimental results shown as MiB/s PQ-throughput, where one item is eight bytes. The PQ implementations are abbreviated as Parallel PQ (PPQ), parallelized STXXL PQ (SPQP), sequential STXXL PQ (SPQS), and STXXL Sorter (Sort).}\label{tab:results-8}
\end{table}

% IMPORT-DATA pqorder plotdata/pqorder.txt

% IMPORT-DATA stats125 plotdata/20150330/i10pc125-24b/*/size-*.txt
% IMPORT-DATA stats129 plotdata/20150330/i10pc129-24b/*/size-*.txt

% IMPORT-DATA stats125b plotdata/20150330/i10pc125-24b/*/bulk-*.txt
% IMPORT-DATA stats129b plotdata/20150330/i10pc129-24b/*/bulk-*.txt

\begin{figure}[p]\centering
  \pgfplotsset{
    plotMini,
    ymin=0,
    legend to name={none},
  }
  \begin{tikzpicture}
    \begin{axis}[
      title={\bf Platform A-Rot},
      xlabel={number of items [$\log_2(n)$]},
      ylabel={\clap{million items per second}},
      ]

      \draw[EMline] (axis cs:29,\pgfkeysvalueof{/pgfplots/ymax}) -- (axis cs:29,\pgfkeysvalueof{/pgfplots/ymin});

      %% MULTIPLOT(pqname)
      %% SELECT LOG(2, num_elements) AS x, MEDIAN(num_elements / time / 1e6) AS y, MULTIPLOT
      %% FROM stats125 s LEFT JOIN pqorder p ON p.name = s.pqname
      %% WHERE op='push-rand-pop' GROUP BY p.order,MULTIPLOT,x ORDER BY p.order,MULTIPLOT,x
      \addplot coordinates { (25.415,28.6194) (26.415,24.926) (27.415,28.2509) (28.415,27.275) (29.415,7.51934) (30.415,7.36308) (31.415,7.2143) (32.415,7.92986) (33.415,7.80561) };
      \addlegendentry{pqname=ppq};
      \addplot coordinates { (25.415,9.39955) (26.415,8.82988) (27.415,7.70476) (28.415,7.73042) (29.415,7.26043) (30.415,7.44224) (31.415,6.50798) (32.415,7.32695) (33.415,6.53582) };
      \addlegendentry{pqname=spq};
      \addplot coordinates { (25.415,4.81481) (26.415,4.66474) (27.415,4.32429) (28.415,3.68128) (29.415,3.64287) (30.415,3.77532) (31.415,3.42147) (32.415,3.80221) (33.415,3.78421) };
      \addlegendentry{pqname=spqs};
      \addplot coordinates { (25.415,8.91689) (26.415,8.92393) (27.415,8.87318) (28.415,8.10519) (29.415,9.14318) (30.415,10.126) (31.415,8.109) (32.415,6.31634) (33.415,11.0096) };
      \addlegendentry{pqname=sorter};

    \end{axis}
  \end{tikzpicture}
  \hfill%
  \begin{tikzpicture}
    \begin{axis}[
      title={\bf Platform B-SSD},
      xlabel={number of items [$\log_2(n)$]},
      ylabel right={\bf push-rand-pop},
      legend to name=mylegend,
      legend columns=2,
      legend transposed=true,
      ]

      \draw[EMline] (axis cs:29,\pgfkeysvalueof{/pgfplots/ymax}) -- (axis cs:29,\pgfkeysvalueof{/pgfplots/ymin});

      %% MULTIPLOT(pqname)
      %% SELECT LOG(2, num_elements) AS x, MEDIAN(num_elements / time / 1e6) AS y, MULTIPLOT
      %% FROM stats129 s LEFT JOIN pqorder p ON p.name = s.pqname
      %% WHERE op='push-rand-pop' GROUP BY p.order,MULTIPLOT,x ORDER BY p.order,MULTIPLOT,x
      \addplot coordinates { (25.415,31.6453) (26.415,23.4568) (27.415,32.2277) (28.415,31.6142) (29.415,18.0246) (30.415,22.5166) (31.415,22.1216) (32.415,16.2667) (33.415,18.6349) };
      \addlegendentry{pqname=ppq};
      \addplot coordinates { (25.415,9.79414) (26.415,8.20542) (27.415,6.23498) (28.415,8.09161) (29.415,7.83767) (30.415,7.61754) (31.415,7.46819) (32.415,6.24916) (33.415,4.87794) };
      \addlegendentry{pqname=spq};
      \addplot coordinates { (25.415,5.73433) (26.415,5.54153) (27.415,5.34139) (28.415,4.434) (29.415,4.49119) (30.415,4.46657) (31.415,4.42438) (32.415,4.56079) (33.415,4.45077) };
      \addlegendentry{pqname=spqs};
      \addplot coordinates { (25.415,16.7204) (26.415,16.6864) (27.415,17.3223) (28.415,17.6349) (29.415,18.3921) (30.415,20.2472) (31.415,20.8651) (32.415,21.172) (33.415,20.6917) };
      \addlegendentry{pqname=sorter};

      \legend{Our PPQ, Parallelized STXXL PQ, Sequential STXXL PQ, STXXL Sorter}

    \end{axis}
  \end{tikzpicture}

  \begin{tikzpicture}
    \begin{axis}[
      xlabel={number of items [$\log_2(n)$]},
      ylabel={\clap{million items per second}},
      ]

      \draw[EMline] (axis cs:29,\pgfkeysvalueof{/pgfplots/ymax}) -- (axis cs:29,\pgfkeysvalueof{/pgfplots/ymin});

      %% MULTIPLOT(pqname)
      %% SELECT LOG(2, num_elements) AS x, MEDIAN(num_elements / time / 1e6) AS y, MULTIPLOT
      %% FROM stats125 s LEFT JOIN pqorder p ON p.name = s.pqname
      %% WHERE op='push-asc-pop' GROUP BY p.order,MULTIPLOT,x ORDER BY p.order,MULTIPLOT,x
      \addplot coordinates { (25.415,43.3171) (26.415,41.5871) (27.415,35.253) (28.415,36.2138) (29.415,9.18603) (30.415,9.2814) (31.415,9.29906) (32.415,9.37357) (33.415,9.4459) };
      \addlegendentry{pqname=ppq};
      \addplot coordinates { (25.415,10.3806) (26.415,10.2244) (27.415,9.76019) (28.415,7.92113) (29.415,7.17861) (30.415,6.68476) (31.415,6.25882) (32.415,6.1489) (33.415,6.01227) };
      \addlegendentry{pqname=spq};
      \addplot coordinates { (25.415,6.72017) (26.415,6.68293) (27.415,6.67523) (28.415,5.25397) (29.415,5.0787) (30.415,4.85251) (31.415,4.58134) (32.415,4.47651) (33.415,4.44167) };
      \addlegendentry{pqname=spqs};
      \addplot coordinates { (25.415,7.44417) (26.415,7.34466) (27.415,7.29936) (28.415,6.05705) (29.415,6.61096) (30.415,7.08642) (31.415,7.37771) (32.415,7.51322) (33.415,7.58006) };
      \addlegendentry{pqname=sorter};

    \end{axis}
  \end{tikzpicture}
  \hfill%
  \begin{tikzpicture}
    \begin{axis}[
      xlabel={number of items [$\log_2(n)$]},
      ylabel right={\bf push-asc-pop},
      legend to name=mylegend,
      legend columns=2,
      legend transposed=true,
      ymax=64,
      ]

      \draw[EMline] (axis cs:29,\pgfkeysvalueof{/pgfplots/ymax}) -- (axis cs:29,\pgfkeysvalueof{/pgfplots/ymin});

      %% MULTIPLOT(pqname)
      %% SELECT LOG(2, num_elements) AS x, MEDIAN(num_elements / time / 1e6) AS y, MULTIPLOT
      %% FROM stats129 s LEFT JOIN pqorder p ON p.name = s.pqname
      %% WHERE op='push-asc-pop' GROUP BY p.order,MULTIPLOT,x ORDER BY p.order,MULTIPLOT,x
      \addplot coordinates { (25.415,55.067) (26.415,53.6432) (27.415,46.6461) (28.415,49.5157) (29.415,29.0392) (30.415,28.9838) (31.415,29.4526) (32.415,27.7826) (33.415,26.8615) };
      \addlegendentry{pqname=ppq};
      \addplot coordinates { (25.415,10.893) (26.415,10.6928) (27.415,9.83913) (28.415,9.23596) (29.415,8.72702) (30.415,8.54184) (31.415,8.30422) (32.415,8.20882) (33.415,7.95906) };
      \addlegendentry{pqname=spq};
      \addplot coordinates { (25.415,5.73632) (26.415,5.83096) (27.415,5.82772) (28.415,5.13589) (29.415,5.02748) (30.415,4.97616) (31.415,4.85883) (32.415,4.84465) (33.415,4.83964) };
      \addlegendentry{pqname=spqs};
      \addplot coordinates { (25.415,20.3667) (26.415,21.703) (27.415,23.0145) (28.415,22.7879) (29.415,24.7926) (30.415,28.619) (31.415,30.5786) (32.415,31.2066) (33.415,30.4999) };
      \addlegendentry{pqname=sorter};

      \legend{Our PPQ, Parallelized STXXL PQ, Sequential STXXL PQ, STXXL Sorter}

    \end{axis}
  \end{tikzpicture}

  \begin{tikzpicture}
    \begin{axis}[
      xlabel={number of items [$\log_2(n)$]},
      ylabel={\clap{million items per second}},
      ]

      \draw[EMline] (axis cs:29,\pgfkeysvalueof{/pgfplots/ymax}) -- (axis cs:29,\pgfkeysvalueof{/pgfplots/ymin});

      %% MULTIPLOT(pqname)
      %% SELECT LOG(2, num_elements) AS x, MEDIAN(num_elements / time / 1e6) AS y, MULTIPLOT
      %% FROM stats125 WHERE op='rbulk-pop-push|cycles' GROUP BY MULTIPLOT,x ORDER BY MULTIPLOT,x
      \addplot coordinates { (25.415,116.496) (26.415,89.2135) (27.415,87.9171) (28.415,95.2669) (29.415,62.4959) (30.415,23.0389) (31.415,23.1875) (32.415,23.3579) (33.415,21.823) };
      \addlegendentry{pqname=ppq};
      \addplot coordinates { (25.415,25.004) (26.415,23.6329) (27.415,21.4634) (28.415,16.8576) (29.415,13.9024) (30.415,14.0481) (31.415,13.7438) (32.415,13.4219) (33.415,12.972) };
      \addlegendentry{pqname=spq};
      \addplot coordinates { (25.415,12.8766) (26.415,12.8345) (27.415,12.6373) (28.415,10.8323) (29.415,9.39343) (30.415,9.40159) (31.415,9.29577) (32.415,9.10618) (33.415,8.9256) };
      \addlegendentry{pqname=spqs};

    \end{axis}
  \end{tikzpicture}
  \hfill%
  \begin{tikzpicture}
    \begin{axis}[
      xlabel={number of items [$\log_2(n)$]},
      ylabel right={\bf asc-rbulk-rewrite},
      ]

      \draw[EMline] (axis cs:29,\pgfkeysvalueof{/pgfplots/ymax}) -- (axis cs:29,\pgfkeysvalueof{/pgfplots/ymin});

      %% MULTIPLOT(pqname)
      %% SELECT LOG(2, num_elements) AS x, MEDIAN(num_elements / time / 1e6) AS y, MULTIPLOT
      %% FROM stats129 WHERE op='rbulk-pop-push|cycles' GROUP BY MULTIPLOT,x ORDER BY MULTIPLOT,x
      \addplot coordinates { (25.415,84.3249) (26.415,59.7567) (27.415,78.2938) (28.415,81.3287) (29.415,119.598) (30.415,87.0694) (31.415,66.1111) (32.415,77.6024) (33.415,76.8174) };
      \addlegendentry{pqname=ppq};
      \addplot coordinates { (25.415,32.0693) (26.415,30.1518) (27.415,23.2721) (28.415,32.2829) (29.415,25.9551) (30.415,26.0058) (31.415,25.5127) (32.415,24.0719) (33.415,21.4558) };
      \addlegendentry{pqname=spq};
      \addplot coordinates { (25.415,33.3853) (26.415,33.7549) (27.415,32.2845) (28.415,22.0891) (29.415,20.9945) (30.415,19.7527) (31.415,19.8784) (32.415,20.3871) (33.415,20.2845) };
      \addlegendentry{pqname=spqs};

    \end{axis}
  \end{tikzpicture}

  \begin{tikzpicture}
    \begin{axis}[
      plotMiniBulkSize,
      xlabel={items in bulk sequences [$\log_2(v)$]},
      ylabel={\clap{million items per second}},
      ]

      %% MULTIPLOT(pqname)
      %% SELECT LOG(2, bulk_size) AS x, MEDIAN(num_elements / time / 1e6) AS y, MULTIPLOT
      %% FROM stats125b s LEFT JOIN pqorder p ON p.name = s.pqname
      %% WHERE op='bulk-pop-push|cycles' GROUP BY p.order,MULTIPLOT,x ORDER BY p.order,MULTIPLOT,x
      \addplot coordinates { (12.2877,12.6001) (13.2877,17.4681) (14.2877,20.9564) (15.2877,22.8865) (16.2877,22.8928) (17.2877,23.5844) (18.2877,22.8576) (19.2877,22.6966) (20.2877,23.0494) (21.2877,21.7243) };
      \addlegendentry{pqname=ppq};
      \addplot coordinates { (12.2877,14.8636) (13.2877,14.9974) (14.2877,14.7407) (15.2877,14.8264) (16.2877,14.7937) (17.2877,14.8712) (18.2877,14.6319) (19.2877,13.1153) (20.2877,13.011) (21.2877,13.0572) };
      \addlegendentry{pqname=spq};
      \addplot coordinates { (12.2877,10.6576) (13.2877,10.427) (14.2877,10.3373) (15.2877,10.488) (16.2877,10.4778) (17.2877,10.4695) (18.2877,10.4671) (19.2877,8.52673) (20.2877,8.50445) (21.2877,8.52541) };
      \addlegendentry{pqname=spqs};

    \end{axis}
  \end{tikzpicture}
  \hfill%
  \begin{tikzpicture}
    \begin{axis}[
      plotMiniBulkSize,
      xlabel={items in bulk sequences [$\log_2(v)$]},
      ylabel right={\bf bulk-rewrite},
      ]

      %% MULTIPLOT(pqname)
      %% SELECT LOG(2, bulk_size) AS x, MEDIAN(num_elements / time / 1e6) AS y, MULTIPLOT
      %% FROM stats129b s LEFT JOIN pqorder p ON p.name = s.pqname
      %% WHERE op='bulk-pop-push|cycles' GROUP BY p.order,MULTIPLOT,x ORDER BY p.order,MULTIPLOT,x
      \addplot coordinates { (12.2877,8.23998) (13.2877,14.5452) (14.2877,24.3509) (15.2877,39.1176) (16.2877,58.2559) (17.2877,77.3525) (18.2877,90.2257) (19.2877,97.7246) (20.2877,100.43) (21.2877,99.8205) (22.2877,92.1888) };
      \addlegendentry{pqname=ppq};
      \addplot coordinates { (12.2877,29.6901) (13.2877,30.1808) (14.2877,30.2879) (15.2877,29.5539) (16.2877,29.7189) (17.2877,29.3563) (18.2877,29.2647) (19.2877,23.5256) (20.2877,23.9527) (21.2877,24.4229) (22.2877,24.9213) };
      \addlegendentry{pqname=spq};
      \addplot coordinates { (12.2877,22.1416) (13.2877,22.7271) (14.2877,22.4508) (15.2877,22.4334) (16.2877,22.5501) (17.2877,21.7949) (18.2877,22.3988) (19.2877,18.9075) (20.2877,20.2924) (21.2877,19.2359) (22.2877,19.9905) };
      \addlegendentry{pqname=spqs};

    \end{axis}
  \end{tikzpicture}

  \centerline{\ref{mylegend}}

  \caption{Experimental results of our four benchmarks with 24-byte items.}\label{fig:results-24}
\end{figure}
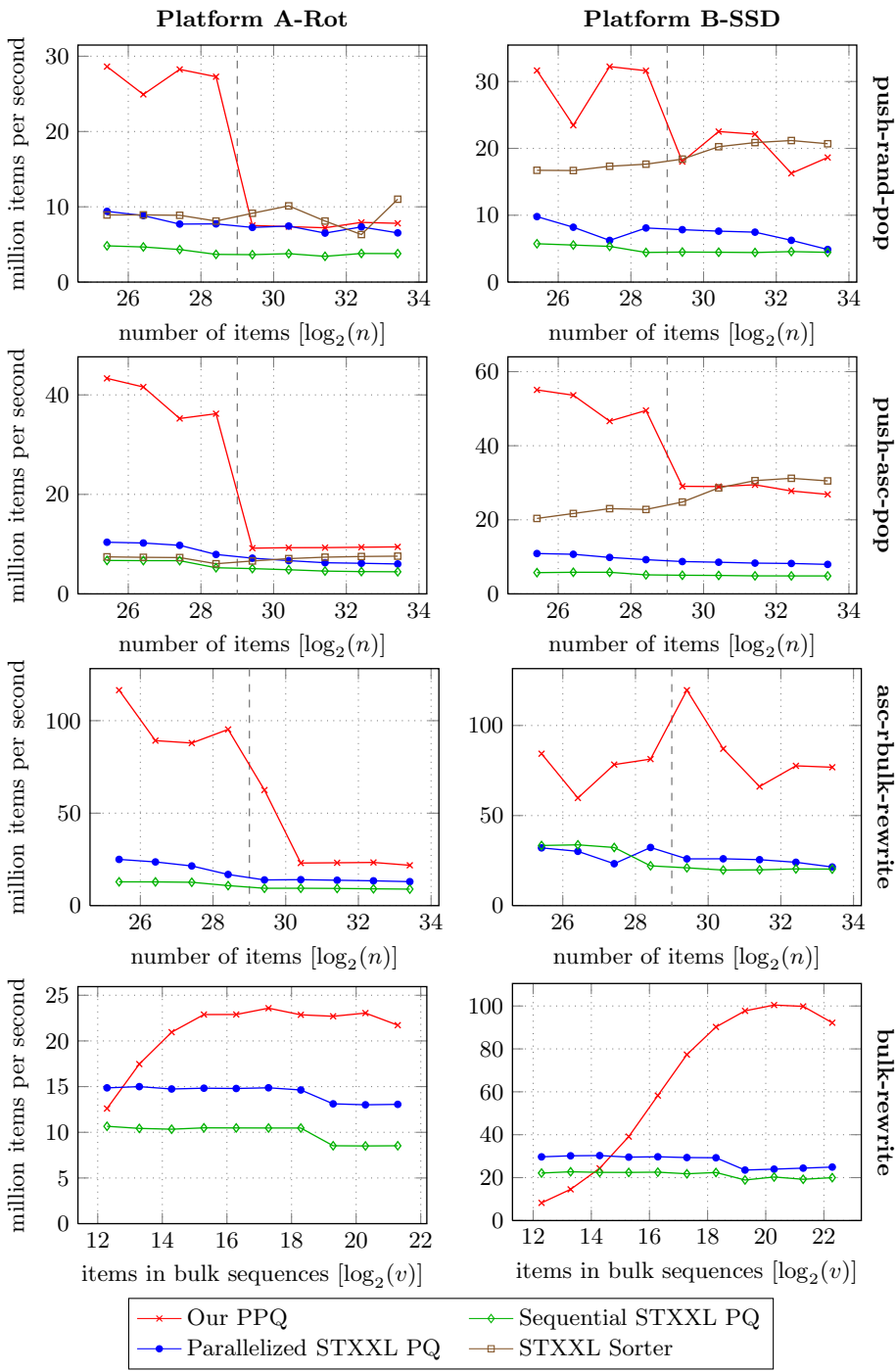

\begin{table}[p]\centering
  \begin{tabular}{|r|r|*4{r}|*4{r}|*3{r}|}
    \hline
    items      & item vol. & \multicolumn{4}{c|}{push-rand-pop} & \multicolumn{4}{c|}{push-asc-pop} & \multicolumn{3}{c|}{asc-rbulk-rewrite}                     \\
    $\log_2 n$ & [GiB]    & PPQ                                & SPQP                              & SPQS & Sort & PPQ & SPQP & SPQS & Sort & PPQ & SPQP & SPQS \\ \hline
    \multicolumn{2}{|c|}{}          & \multicolumn{11}{c|}{Platform A-Rot}                                                                                                \\
    \hline
    25.4 &   0 & 1\,310 & 430 & 220 & 408 & 1\,983 & 475 & 308 & 341 & 5\,333 & 1\,145 & 589 \\
    26.4 &   1 & 1\,141 & 404 & 214 & 409 & 1\,904 & 468 & 306 & 336 & 4\,084 & 1\,082 & 588 \\
    27.4 &   3 & 1\,293 & 353 & 198 & 406 & 1\,614 & 447 & 306 & 334 & 4\,025 &    983 & 578 \\
    28.4 &   7 & 1\,249 & 354 & 169 & 371 & 1\,658 & 363 & 241 & 277 & 4\,361 &    772 & 496 \\ \hdashline
    29.4 &  15 &    344 & 332 & 167 & 419 &    421 & 329 & 232 & 303 & 2\,861 &    636 & 430 \\
    30.4 &  31 &    337 & 341 & 173 & 464 &    425 & 306 & 222 & 324 & 1\,055 &    643 & 430 \\
    31.4 &  63 &    330 & 298 & 157 & 371 &    426 & 287 & 210 & 338 & 1\,061 &    629 & 426 \\
    32.4 & 127 &    363 & 335 & 174 & 289 &    429 & 281 & 205 & 344 & 1\,069 &    614 & 417 \\
    33.4 & 255 &    357 & 299 & 173 & 504 &    432 & 275 & 203 & 347 &    999 &    594 & 409 \\
    % END TABULAR SELECT LOG(2, num_elements) AS logsize, num_elements * value_si...
    \hline
    \multicolumn{2}{|c|}{} & \multicolumn{11}{c|}{Platform B-SSD}                                       \\
    \hline
    25.4 &   0 & 1\,449 & 448 & 262 & 765 & 2\,521 & 499 & 263 &    932 & 3\,860 & 1\,468 & 1\,528 \\
    26.4 &   1 & 1\,074 & 376 & 254 & 764 & 2\,456 & 489 & 267 &    993 & 2\,735 & 1\,380 & 1\,545 \\
    27.4 &   3 & 1\,475 & 285 & 245 & 793 & 2\,135 & 450 & 267 & 1\,054 & 3\,584 & 1\,065 & 1\,478 \\
    28.4 &   7 & 1\,447 & 370 & 203 & 807 & 2\,267 & 423 & 235 & 1\,043 & 3\,723 & 1\,478 & 1\,011 \\ \hdashline
    29.4 &  15 &    825 & 359 & 206 & 842 & 1\,329 & 399 & 230 & 1\,135 & 5\,475 & 1\,188 &    961 \\
    30.4 &  31 & 1\,031 & 349 & 204 & 927 & 1\,327 & 391 & 228 & 1\,310 & 3\,986 & 1\,190 &    904 \\
    31.4 &  63 & 1\,013 & 342 & 203 & 955 & 1\,348 & 380 & 222 & 1\,400 & 3\,026 & 1\,168 &    910 \\
    32.4 & 127 &    745 & 286 & 209 & 969 & 1\,272 & 376 & 222 & 1\,429 & 3\,552 & 1\,102 &    933 \\
    33.4 & 255 &    853 & 223 & 204 & 947 & 1\,230 & 364 & 222 & 1\,396 & 3\,516 &    982 &    929 \\
    % END TABULAR SELECT LOG(2, num_elements), num_elements * value_size / 1024 /...
    \hline
  \end{tabular}
  \medskip

  \def\tabcolsep{6pt}
  \begin{tabular}{|r|r|*3{r}|*3{r}|}
    \hline
    bulk size  & bulk vol. & \multicolumn{3}{c|}{A-Rot} & \multicolumn{3}{c|}{B-SSD}      \\
    $n / 1000$ & [KiB]     & PPQ                        & SPQP & SPQS & PPQ & SPQP & SPQS \\
    \hline
    %% TABULAR REFORMAT(precision=0 group=\,)
    %% SELECT bulk_size / 1000, bulk_size * value_size / 1024,
    %% (SELECT MEDIAN( num_elements * value_size * 2 / time / 1024 / 1024 ) FROM stats125b b WHERE b.pqname='ppq' AND op='bulk-pop-push|cycles' AND b.bulk_size = a.bulk_size),
    %% (SELECT MEDIAN( num_elements * value_size * 2 / time / 1024 / 1024 ) FROM stats125b b WHERE b.pqname='spq' AND op='bulk-pop-push|cycles' AND b.bulk_size = a.bulk_size),
    %% (SELECT MEDIAN( num_elements * value_size * 2 / time / 1024 / 1024 ) FROM stats125b b WHERE b.pqname='spqs' AND op='bulk-pop-push|cycles' AND b.bulk_size = a.bulk_size),
    %% (SELECT MEDIAN( num_elements * value_size * 2 / time / 1024 / 1024 ) FROM stats129b b WHERE b.pqname='ppq' AND op='bulk-pop-push|cycles' AND b.bulk_size = a.bulk_size),
    %% (SELECT MEDIAN( num_elements * value_size * 2 / time / 1024 / 1024 ) FROM stats129b b WHERE b.pqname='spq' AND op='bulk-pop-push|cycles' AND b.bulk_size = a.bulk_size),
    %% (SELECT MEDIAN( num_elements * value_size * 2 / time / 1024 / 1024 ) FROM stats129b b WHERE b.pqname='spqs' AND op='bulk-pop-push|cycles' AND b.bulk_size = a.bulk_size)
    %% FROM stats125b a GROUP BY bulk_size, num_elements, value_size ORDER BY bulk_size
         5 &     117 &    577 & 680 & 488 &    377 & 1\,359 & 1\,014 \\
        10 &     234 &    800 & 687 & 477 &    666 & 1\,382 & 1\,040 \\
        20 &     468 &    959 & 675 & 473 & 1\,115 & 1\,386 & 1\,028 \\
        40 &     937 & 1\,048 & 679 & 480 & 1\,791 & 1\,353 & 1\,027 \\
        80 &  1\,875 & 1\,048 & 677 & 480 & 2\,667 & 1\,360 & 1\,032 \\
       160 &  3\,750 & 1\,080 & 681 & 479 & 3\,541 & 1\,344 &    998 \\
       320 &  7\,500 & 1\,046 & 670 & 479 & 4\,130 & 1\,340 & 1\,025 \\
       640 & 15\,000 & 1\,039 & 600 & 390 & 4\,473 & 1\,077 &    866 \\
    1\,280 & 30\,000 & 1\,055 & 596 & 389 & 4\,597 & 1\,096 &    929 \\
    2\,560 & 60\,000 &    994 & 598 & 390 & 4\,569 & 1\,118 &    881 \\
    % END TABULAR SELECT bulk_size / 1000, bulk_size * value_size / 1024, (SELECT...
    \hline
  \end{tabular}
  \medskip
  \caption{Experimental results shown as MiB/s PQ-throughput, where one item is 24~bytes. The PQ implementations are abbreviated as Parallel PQ (PPQ), parallelized STXXL PQ (SPQP), sequential STXXL PQ (SPQS), and STXXL Sorter (Sort).}\label{tab:results-24}
\end{table}

\begin{table}[p]\centering
  \def\tabcolsep{6pt}
  \begin{tabular}{|r|rrr|rrr|}
    \hline
               & \multicolumn{3}{c|}{Platform A-Rot} & \multicolumn{3}{c|}{Platform B-SSD}  \\
    Experiment & SPQP                                & SPQS & Sorter & SPQP & SPQS & Sorter \\ \hline
    %% TABULAR REFORMAT(precision=2)
    %% SELECT opo.desc,
    %% MEDIAN(spq125.time / ppq125.time) AS spq_ratio,
    %% MEDIAN(spqs125.time / ppq125.time) AS spqs_ratio,
    %% MEDIAN(sort125.time / ppq125.time) AS sort_ratio,
    %% MEDIAN(spq129.time / ppq129.time) AS spq_ratio,
    %% MEDIAN(spqs129.time / ppq129.time) AS spqs_ratio,
    %% MEDIAN(sort129.time / ppq129.time) AS sort_ratio
    %% FROM stats125 s
    %% LEFT JOIN oporder opo ON opo.name = s.op
    %% LEFT JOIN stats125 ppq125  ON ppq125.pqname='ppq'     AND ppq125.op = s.op  AND ppq125.num_elements = s.num_elements
    %% LEFT JOIN stats125 spq125  ON spq125.pqname='spq'     AND spq125.op = s.op  AND spq125.num_elements = s.num_elements
    %% LEFT JOIN stats125 spqs125 ON spqs125.pqname='spqs'   AND spqs125.op = s.op AND spqs125.num_elements = s.num_elements
    %% LEFT JOIN stats125 sort125 ON sort125.pqname='sorter' AND sort125.op = s.op AND sort125.num_elements = s.num_elements
    %% LEFT JOIN stats129 ppq129  ON ppq129.pqname='ppq'     AND ppq129.op = s.op  AND ppq129.num_elements = s.num_elements
    %% LEFT JOIN stats129 spq129  ON spq129.pqname='spq'     AND spq129.op = s.op  AND spq129.num_elements = s.num_elements
    %% LEFT JOIN stats129 spqs129 ON spqs129.pqname='spqs'   AND spqs129.op = s.op AND spqs129.num_elements = s.num_elements
    %% LEFT JOIN stats129 sort129 ON sort129.pqname='sorter' AND sort129.op = s.op AND sort129.num_elements = s.num_elements
    %% WHERE s.pqname='ppq' AND s.op IN ('push-rand-pop', 'push-asc-pop', 'rbulk-pop-push|cycles')
    %% AND LOG(2, s.num_elements) >= 30.5
    %% GROUP BY opo.order, opo.desc
    %% ORDER BY opo.order
        push-rand-pop & 1.11 & 2.09 & 0.89 & 3.00 & 4.19 & 0.90 \\
         push-asc-pop & 1.52 & 2.09 & 1.25 & 3.39 & 5.73 & 0.89 \\
    asc-rbulk-rewrite & 1.69 & 2.49 &      & 3.22 & 3.72 &      \\
    % END TABULAR SELECT opo.desc, MEDIAN(spq125.time / ppq125.time) AS spq_ratio...
    \hline
  \end{tabular}
  \vspace{2ex}
  \caption{Speedup of PPQ over parallelized STXXL PQ, sequential STXXL PQ, and STXXL Sorter for 24~byte structures, averaged for all experiments with $n \geq 2^{30.5}$.}\label{tab:speedups-24}
\end{table}

\end{document}